\documentclass[trackchanges,twocolumn]{aastex701}

\newcommand{\ca}[1]{\textcolor{black}{#1}} %{\textbf{#1}}}

\newcommand{\cb}[1]{\textcolor{black}{#1}} %{\textbf{#1}}}

\newcommand{\cc}[1]{\textcolor{black}{#1}} %{\textbf{#1}}}

\newcommand{\cd}[1]{\textcolor{black}{#1}} %{\textbf{#1}}}

\newcommand{\ce}[1]{\textcolor{black}{#1}} %{\textbf{#1}}}

\newcommand{\cf}[1]{\textcolor{black}{#1}} %{\textbf{#1}}}

\newcommand{\cg}[1]{\textcolor{black}{#1}} %{\textbf{#1}}}

\newcommand{\ch}[1]{\textcolor{black}{#1}} %{\textbf{#1}}}

\newcommand{\ckt}[1]{\textcolor{black}{#1}} 

\newcommand{\ci}[1]{\textcolor{black}{#1}}

\newcommand{\cj}[1]{\textcolor{black}{#1}}

\newcommand{\ck}[1]{\textcolor{black}{#1}}

\newcommand{\cl}[1]{\textcolor{black}{#1}}

\newcommand{\cn}[1]{\textcolor{black}{#1}}

\newcommand{\co}[1]{\textcolor{black}{#1}}
\newcommand{\cp}[1]{\textcolor{black}{#1}}
\newcommand{\cq}[1]{\textcolor{black}{#1}}

\newcommand{\NAOJ}{National Astronomical Observatory of Japan, 2-21-1 Osawa, Mitaka, Tokyo 181-8588, Japan}
\newcommand{\SOKENDAI}{Department of Astronomical Science, The Graduate University for Advanced Studies, SOKENDAI, 2-21-1 Osawa, Mitaka, Tokyo 181-8588, Japan}

\newcommand{\ICRR}{Institute for Cosmic Ray Research, The University of Tokyo, 5-1-5 Kashiwanoha, Kashiwa, Chiba 277-8582, Japan}
\newcommand{\UT}{Department of Physics, Graduate School of Science, The University of Tokyo, 7-3-1 Hongo, Bunkyo, Tokyo 113-0033, Japan}
\newcommand{\TSUKUBA}{Center for Computational Sciences, University of Tsukuba, Ten-nodai, 1-1-1 Tsukuba, Ibaraki 305-8577, Japan}

\usepackage{lmodern}
\usepackage{comment}
\usepackage{amsmath}
\usepackage{url}

\begin{document}

\title{
DREAMS.\\
\textit{JWST} Spectroscopy of a $z=8.3$ Galaxy with an ALMA Dust Continuum Detection:\\
Early Dust, Very High $T_{\rm dust}$, and a Multi-wavelength [\ion{O}{3}] Ratio Discrepancy\\
}

\author[0009-0008-1353-7823]{Kana Takechi}
\email[show]{ktakechi@icrr.u-tokyo.ac.jp }  
\affiliation{\ICRR}
\affiliation{\UT}

%\begin{comment}

\author[0000-0002-1049-6658]{Masami Ouchi}
\affiliation{\ICRR}
\affiliation{\NAOJ}
\affiliation{Astronomical Science Program, Graduate Institute for Advanced Studies, SOKENDAI, 2-21-1 Osawa, Mitaka, Tokyo 181-8588, Japan}
\affiliation{Kavli Institute for the Physics and Mathematics of the Universe (Kavli IPMU, WPI), The University of Tokyo, 5-1-5 Kashiwanoha, Kashiwa, Chiba, 277-8583, Japan}
\email{ouchims@icrr.u-tokyo.ac.jp}

\author[0000-0003-2965-5070]{Kimihiko Nakajima}
\affiliation{Institute of Liberal Arts and Science, Kanazawa University, Kakuma-machi, Kanazawa, Ishikawa, 920-1192, Japan}
\affiliation{\NAOJ}
\email{knakajima@staff.kanazawa-u.ac.jp}

\author[0009-0004-4332-9225]{Tomokazu Kiyota}
\affiliation{\NAOJ}
\affiliation{\SOKENDAI}
\email{tomokazu.kiyota@grad.nao.ac.jp}

\author[0000-0003-4807-8117]{Yoichi Tamura}
\affiliation{Department of Physics, Graduate School of Science, Nagoya University, Furo, Chikusa, Nagoya, Aichi 464-8602, Japan}
\email{ytamura@nagoya-u.jp}

\author[0000-0002-6047-430X]{Yuichi Harikane}
\affiliation{\ICRR} 
\email{hari@icrr.u-tokyo.ac.jp}

\author[0000-0002-0984-7713]{Yurina Nakazato}
\affiliation{Center for Computational Astrophysics, Flatiron Institute, 162 5th Avenue, New York, NY 10010}
\email{ynakazato@flatironinstitute.org}

\author[0000-0002-5268-2221]{Tom J.\ L.\ C.\ Bakx}
\affiliation{Department of Physics and Astronomy, Chalmers University of Technology, SE-412 96 Gothenburg, Sweden}
\email{tjlcbakx@gmail.com}

\author[0000-0002-7779-8677]{Akio K.\ Inoue}
\affiliation{Department of Pure and Applied Physics, Graduate School of Advanced Science and Engineering, Faculty of Science and Engineering, Waseda University, 3-4-1, Okubo, Shinjuku, Tokyo 169-8555, Japan}
\affiliation{Waseda Research Institute for Science and Engineering, Faculty of Science and Engineering, Waseda University, \\3-4-1, Okubo, Shinjuku, Tokyo 169-8555, Japan}
\email{akinoue@aoni.waseda.jp}

\author[0000-0002-1319-3433]{Hidenobu Yajima}
\affiliation{\TSUKUBA} 
\email{yajima@ccs.tsukuba.ac.jp}

\author[0000-0001-8083-5814]{Masato Hagimoto}
\affiliation{Department of Physics, Graduate School of Science, Nagoya University, Furo, Chikusa, Nagoya, Aichi 464-8602, Japan}
\email{hagimoto@a.phys.nagoya-u.ac.jp}

\author[0000-0001-9011-7605]{Yoshiaki Ono}
\affiliation{\ICRR}
\email{ono@icrr.u-tokyo.ac.jp}

\author[0000-0002-5768-8235]{Yi Xu}
\affiliation{Cosmic Dawn Center (DAWN), Niels Bohr Institute, University of Copenhagen, Jagtvej 128, Copenhagen N, DK-2200, Denmark.}
\email{yi.xu@nbi.ku.dk}

\begin{abstract}

%246 words
We present a deep DREAMS \textit{JWST}/NIRSpec MSA medium-grating spectrum of MACS0416-Y1, a galaxy at $z=8.312$ with the highest-redshift ALMA dust continuum detection to date, in order to characterize its properties together with archival IFU and ALMA data. The deep NIRSpec spectrum reveals a broad H$\beta$ line with a width of $\sim1100$ km~s$^{-1}$. We interpret it as a broad-line AGN whose line diagnostics are consistent with AGN activity across its clumpy structure, given the absence of little red dot signatures.
MACS0416-Y1 clearly shows [\textsc{Oiii}]\,4363 emission, suggesting a moderately low metallicity of $12+\log(\mathrm{O/H})=7.86^{+0.09}_{-0.08}$~($0.15~Z_\odot$). The combination of [\textsc{Cii}]\,158$\mu$m and dust continuum emission indicates low dust mass ratios of $\log (M_{\rm dust}/M_{\rm gas})=-3.60^{+0.29}_{-0.22}$ and $\log (M_{\rm dust}/M_{\rm metal})=-0.95^{+0.29}_{-0.20}$. Because the metallicity of MACS0416-Y1 is around the critical metallicity of $0.1\textrm{--}0.2~Z_\odot$, the system is expected to undergo dust growth, explaining these low dust mass ratios as well as its small dust mass, $M_{\rm dust}\sim10^6~M_\odot$. The intense UV radiation from the AGN may contribute to a high dust temperature of $T_{\rm dust}\simeq91^{+62}_{-35}$ K, boosting the dust-continuum emission above the ALMA detection limit despite the small $M_{\rm dust}$ at $z>8$. We find a very high total flux ratio of [\textsc{Oiii}]\,88$\mu$m/[\textsc{Oiii}]\,5007 = $0.26 \pm 0.06$ in MACS0416-Y1, above predictions from single ionized nebular models at any electron density. This discrepancy suggests that the [\textsc{Oiii}]\,88$\mu$m and [\textsc{Oiii}]\,5007 trace largely distinct regions, with the optical line suppressed in dusty nebulae, and thus requires careful interpretation when combining optical and infrared emission lines in JWST+ALMA studies.

\end{abstract}

\section{Introduction}
Dust is a fundamental component of the interstellar medium (ISM) in galaxies, influencing the observational properties, energy budget, and physical conditions. In the local \cd{Universe}, dust heated by interstellar radiation fields typically reaches equilibrium temperatures of $\sim$20--40 K~\ca{\citep{Dunne2000,Casey2012}}. However, recent far-infrared (FIR) and submillimeter observations have revealed that dust in high-redshift galaxies, particularly at $z > 6$, exhibits unexpectedly high temperatures of 50--100 K, \cc{likely due to intense radiation fields associated with compact star formation and/or conditions that enhance dust heating, including low dust mass or active galactic nucleus (AGN) activity} \citep{Nakazato2026,Viero2022,Sommovigo2022,Mitsuhashi2024,Tsukui2023}.

\cc{Understanding the origin of these high dust temperatures and the physical conditions of the ISM requires combining dust continuum measurements with constraints on the gas properties.}
A multi-wavelength approach that combines FIR/submillimeter continuum observations with near-infrared spectroscopy is therefore essential for characterizing dust and ISM properties in early galaxies. FIR and submillimeter observations probe the emission from dust and \ckt{FIR} fine-structure emission lines, while near-infrared spectroscopy provides emission line diagnostics that trace gas properties including electron temperature, density, ionization state, and chemical composition~\cc{\citep{Osterbrock2006,Draine2011}.} The combination of these datasets enables a more comprehensive understanding of the physical conditions in high-$z$ galaxies. For example, measurements of the gas-phase metallicity \cd{can} provide constraints on grain growth processes in the early universe~\citep[\cc{e.g.,}][]{Asano2013, Popping2017,Toyouchi2025}.

MACS J0416-Y1 (hereafter Y1) is a star-forming galaxy at $z = 8.31$ that exhibits an elevated dust temperature and a complex ISM structure~\ca{characterized by three emission peaks in the rest-frame UV and two emission peaks in the rest-frame FIR continuum}. It was originally identified as a $z>8$ Lyman break galaxy candidate based on multiband \textit{Hubble Space Telescope} Wide Field Camera 3 (\textit{HST}/WFC3) observations~\citep{Infante2015, Laporte2015} in the Hubble Frontier Fields \citep[HFF,][]{Lotz2017} survey of the MACS J-0416.1-2403 cluster. Follow up observations \cn{with} \cl{the Atacama Large Millimeter/submillimeter Array (ALMA) }spectroscopically confirmed its redshift, $z=8.312$, through detections of the [\textsc{Oiii}]\,88$\mu$m~\citep{Tamura2019} and [\textsc{Cii}]\,158$\mu$m~\citep{Bakx2020} emission lines. \co{Y1 exhibits a high [\textsc{Oiii}]\,88$\mu$m/[\textsc{Cii}]\,158$\mu$m luminosity ratio of $\sim9$ \citep{Bakx2020}. \citet{Hagimoto2025} suggested that this can be explained by a porous ISM geometry with a low neutral-gas covering fraction.}
\cg{Dust continuum emission was first reported in ALMA Band 7~\citep{Tamura2019}, representing the highest-redshift dust continuum detection to date.} %This detection implies rapid dust production and enrichment within the first $\sim 600$ Myr after the Big Bang. 
Dust continuum emission has subsequently been detected across multiple ALMA bands \citep{Tamura2019,Harshan2024,Bakx2025}. By combining these detections with upper limits from nondetections~\citep{Bakx2020, Jones2024,Bakx2025}, the dust temperature has been constrained to 91$^{+62}_{-35}$~K~\citep{Bakx2025}, significantly \cd{warmer} than typical values observed in local galaxies~\citep[$\sim25$~K; e.g.,][]{Schreiber2018}. The origin of the high dust temperature and the detection of dust continuum at such a high redshift remain under debate. \citet{Tamura2019} suggested that the observed dust mass requires an earlier episode of star formation to pre-enrich the system, rather than production in a single young burst. \citet{Bakx2020} argue that the high inferred dust temperature itself reduces the required dust mass, alleviating the tension with rapid dust production. 
\cc{More recently, observations with \cl{the James Webb Space Telescope/Near Infrared Spectrograph (\textit{JWST}/NIRSpec) and \textit{JWST}/Near Infrared Camera (NIRCam)} have revealed the clumpy morphology, ionized gas properties, and stellar populations~\citep[e.g.,][]{Ma2024,Harshan2024}.} 
\cp{\citet{Harshan2024} proposed that merger-driven star formation and/or a possible AGN contribution inferred from emission line diagnostics could account for the strong dust heating.}

\ckt{Despite these advances, several key questions remain unresolved, \cn{including the origin of the dust heating, the mechanism responsible for the dust enrichment, and the connection between the rest-frame UV/optical-emitting regions observed with \textit{JWST} and the rest-frame FIR-emitting regions observed with ALMA. Addressing these questions requires a multiwavelength analysis of the ionized gas, cold gas, and dust components using both \textit{JWST} and ALMA observations.} In this study, we present new deep \textit{JWST}/NIRSpec \cg{Micro-Shutter Assembly (MSA)} spectroscopy of Y1, combined with archival \cg{NIRSpec Integral Field Unit (IFU)} and ALMA data, to investigate its ISM properties and the mechanism of dust production and heating.}
\cc{The remainder of this paper is organized as follows. Section \ref{sec:2} describes the observations and datasets used in this study. Section \ref{sec:3} outlines the data reduction and analysis methods. In Section \ref{sec:4}, we present the main results, including emission-line properties and ISM diagnostics. Section \ref{sec:5} discusses the implications for dust heating, production, and ISM structure. Finally, Section \ref{sec:6} summarizes our conclusions.} \cl{Throughout this paper, we assume a flat $\Lambda$CDM cosmology with $H_0=67.66$~km~s$^{-1}$~Mpc$^{-1}$, $\Omega_{\rm m}=0.3111$, and $\Omega_{\Lambda}=0.6889$~\citep{Planck2020}. }

\section{Observations and Data}
\label{sec:2}
\subsection{Observations}

We \cd{have} observed MACS J0416-Y1 (Y1) with \textit{JWST}/NIRSpec \cl{MSA} on 2024 November 5, as part of the General Observers (GO) program Deep Reconnaissance of Early Assemblies of Metal-poor Star formation (DREAMS; proposal ID: 4750; Nakajima et al. in preparation). 
\cd{These} observations were primarily designed to obtain deep NIRSpec MSA spectroscopy of LAP1-B in the MACS J0416 field~\ca{\citep{Nakajima2025}}. Y1, located close to LAP1-B within the NIRSpec MSA footprint, was simultaneously observed together with other high-redshift galaxies in the same field. The observations were carried out using the F290LP/G395M filter/grating configuration, covering 2.87--5.10$~\mu$m at a spectral resolution of $R\sim 1000$. The total effective exposure time \cd{was} 23 ks. 
%\co{We simultaneously observed other high redshift galaxies in the same field, including Y1 \cd{close to} LAP1-B for the slit spectroscopy \cd{within} the NIRSpec MSA footprint.}
%The filter/grating used in these observations is F290LP/G395\cb{\ca{M}}, corresponding to the wavelength coverage of 2.87--5.10$~\mu$m and the spectral resolution of $R\sim 1000$.

\subsection{Archival data}
\subsubsection{\textit{JWST}/NIRSpec IFU}
We use archival \textit{JWST}/NIRSpec IFU data obtained as part of the Guaranteed Time Observation program \# GTO 1208 \citep[CANUCS;][]{Willott2022} on 2023 February 13. The observations were carried out using the F290LP/G395H filter/grating configuration, covering 2.87--5.14~$\mu$m at a spectral resolution of $R\sim2700$. \ca{The observations were conducted with eight dithers and a medium cycling pattern.} The total effective exposure time \cd{was} 18 ks.

%The filter/grating used in these observations is F290LP/G395H, corresponding to the wavelength coverage of 2.87--5.14$~\mu$m and the spectral resolution of $R\sim2700$. \ca{The observations were conducted with eight dithers and a medium cycling pattern.} The total effective exposure time \cd{was} 18 ks.
%The observations were carried out using the F290LP/G395H filter/grating configuration, providing wavelength coverage of 2.87--5.14$~\mu$m and the spectral resolution of $R\sim 2700$. 

\subsubsection{\textit{JWST}/NIRCam Imaging}
\ci{We use archival \textit{JWST}/NIRCam F150W and F444W images of the MACS0416 field observed as part of the Guaranteed Time Observation programs \# GTO 1176 \citep[PEARLS;][]{Windhorst2023} and \# GTO 1208 \citep[CANUCS;][]{Willott2022}. The total integration times are 15 ks and 18 ks for F150W and F444W images, respectively.}
\cj{Y1 lies behind the Hubble Frontier Fields cluster MACS J0416.1$-$2403 and is moderately magnified by gravitational lensing, \cn{with a magnification factor of $\mu =1.60^{+0.01}_{-0.02}$~\citep{Harshan2024}, derived using the lens model presented by \citep{Rihtarsic2024}.} Unless otherwise stated, we do not apply corrections for lensing magnification in this work. }

\subsubsection{ALMA}
We use archival ALMA Band 7 data with project IDs \#2016.1.00117.S \citep[PI: Y. Tamura;][]{Tamura2019}, \#2017.1.0025.S \citep[PI: Y. Tamura;][]{Tamura2023}, \#2017.1.00486.S (PI: R. Ellis), and \#2018.1.01241.S \citep[PI: Y. Tamura;][]{Tamura2023}, observed in Cycles 4--6 from 2016 October to 2019 August \cd{at frequencies around $\sim$353~GHz }\citep[see Table 1 of][]{Tamura2023}. \cn{The total on-source integration times are 27.52 and 18.89 hours for the continuum and [\textsc{Oiii}]\,88$\mu$m observations, respectively.} The baseline lengths range from 15 to 3637~m.
\ckt{We independently re-reduced and analyzed the data and confirmed consistency with the results presented by \citet{Tamura2023}. We therefore adopt their dataset in this work. }

\section{Data Reduction and Analysis}
\label{sec:3}
\subsection{Data Reduction}
\subsubsection{\textit{JWST}/NIRSpec MSA}
We reduced the data using the \textit{JWST} Science Calibration Pipeline (version 1.17.1) with the Calibration Reference Data System (CRDS) context file \texttt{jwst\_1298.pmap}. A detailed description of the reduction procedure will be presented in Nakajima et al. (in preparation; see also \citealt{Nakajima2025}).

\subsubsection{\textit{JWST}/NIRSpec IFU}
We reduced the raw data retrieved from the Mikulski Archive for Space Telescopes (MAST) portal using the \textit{JWST} Science Calibration Pipeline with the CRDS context file \texttt{jwst\_1364.pmap}. We follow the NIRSpec/IFU Data Reduction Pipeline developed by the \textit{JWST} \cl{Early Release Science (ERS)} Targeting Extremely Magnified Panchromatic Lensed Arcs and their Extended Star formation~\ca{\citep[TEMPLATES;][]{Rigby2025}} team through stage 1 and stage 2, and replace stage 3 with custom processing using the \texttt{reproject} and \texttt{reproject\_interp} packages. In stage 1, \texttt{calwebb\_detector1} is applied to perform detector-level corrections for each exposure, followed by $1/f$ noise removal using the \texttt{NSClean} package. In stage 2, \texttt{calwebb\_spec2} is used for WCS assignment, wavelength and flux calibration, flat-fielding, and pass-loss correction. The calibrated data are then combined using drizzle weighting to achieve a spaxel scale of 0$\farcs$05. Finally, we construct the median-stacked 3D data cube using the \texttt{reproject} and \texttt{reproject\_interp} packages, with the background modeled and subtracted as a polynomial function of wavelength. 

\subsubsection{\textit{JWST}/NIRCam Imaging}
\ci{\textit{JWST}/NIRCam raw imaging data were retrieved from MAST archive and reduced using the \textit{JWST} Calibration Pipeline (version 1.12.5) and the CRDS context file \texttt{jwst\_1193.pmap}, with custom modifications following \citet{Harikane2023a}. The final images are drizzled to a pixel scale of $0\farcs015$.}

\subsubsection{ALMA}
We calibrated and imaged the ALMA raw data, retrieved from the ALMA Science Archive using Common Astronomy Software Applications \citep[CASA; ][]{CASA2022}. Data for each program are calibrated using the corresponding CASA pipelines versions: 4.7.0, 5.4.0-68, 5.1.1, and 5.4.0-70 for \#2016.1.00117.S, \#2017.1.0025.S, \#2017.1.00486.S, and \#2018.1.01241.S, respectively. All calibrated measurement sets are concatenated into one measurement set using CASA task \texttt{concat}. The concatenated data are then imaged with the CASA task \texttt{tclean}. \cd{Data from spectral windows (spw) covering 364.1--364.7~GHz were used for [\textsc{Oiii}]\,88$\mu$m line cube imaging with a velocity resolution of 12.85~km~s$^{-1}$}, and all remaining data are used to create the continuum image~\ca{with a pixel size of 0$\farcs$01~pixel$^{-1}$}. We adopted natural weighting, resulting in synthesized beam sizes of 81.1$\times$112.2~mas and 73.9$\times$96.1~mas with position angles of 89$^{\circ}$.4 and 93$^{\circ}$.8 for the line and continuum images, respectively.

\subsection{Astrometric Alignment}
\cb{We calibrated the astrometry of the \textit{JWST} NIRCam and \textit{JWST} NIRSpec/IFU data using an \cj{\textit{HST} image aligned to the \textit{Gaia} DR3~\citep{Gaia2021}} as the astrometric reference.} \cl{For the \textit{HST} image, we used the public \textit{HST} Frontier Fields Epoch 2 data release for MACSJ0416.1$-$2403 \cn{\citep[Proposal ID: 13496; PI: J. Lotz;][]{Lotz2017}}, obtained from MAST.}
To increase the number of \textit{Gaia} sources for astrometric alignment, we utilized the \textit{HST} F125W image to transfer the astrometry from \textit{Gaia} to \textit{JWST}. The \textit{HST} image has a wider field of view and contains fewer saturated sources than the \textit{JWST} images. The \textit{HST} astrometry was first aligned to \textit{Gaia} DR3 catalog coordinates using 4 \textit{Gaia} sources within the field of view. The NIRCam images were then aligned to the \textit{HST} image using 4 stars detected in both datasets. 
We constructed a pseudo F444W image of Y1 from NIRSpec/IFU data and aligned it to the NIRCam F444W image. \co{After astrometric calibration, the mean positional offset among the matched reference sources is 10 mas.}

\subsection{Line Flux and Profile Measurements}
\cn{We fit the emission lines in the NIRSpec/MSA and IFU spectra using a Markov Chain Monte Carlo (MCMC) method with \texttt{emcee}~\citep{ForemanMackey2013}.} Flux uncertainties correspond to the 16th and 84th percentiles of the posterior distributions. \cn{The flux-density uncertainties \cn{were} estimated as $1\sigma$ of the flux density in nearby line-free spectral windows within the same extraction region.}
The emission-line FWHMs \cn{were} constrained to be larger than the FWHM of the instrumental line spread function (LSF) measured from planetary nebula spectra in \citet{Isobe2023}. 
\cn{The intrinsic emission-line velocity widths presented in this paper \cn{were} derived by deconvolving a Gaussian profile whose FWHM matches that of the LSF measured by \citet{Isobe2023}.}

\section{Results}
\label{sec:4}

\subsection{Morphology, Spectrum, and Dynamics}
\begin{figure}[h]
    \centering
    \includegraphics[width=1\linewidth]{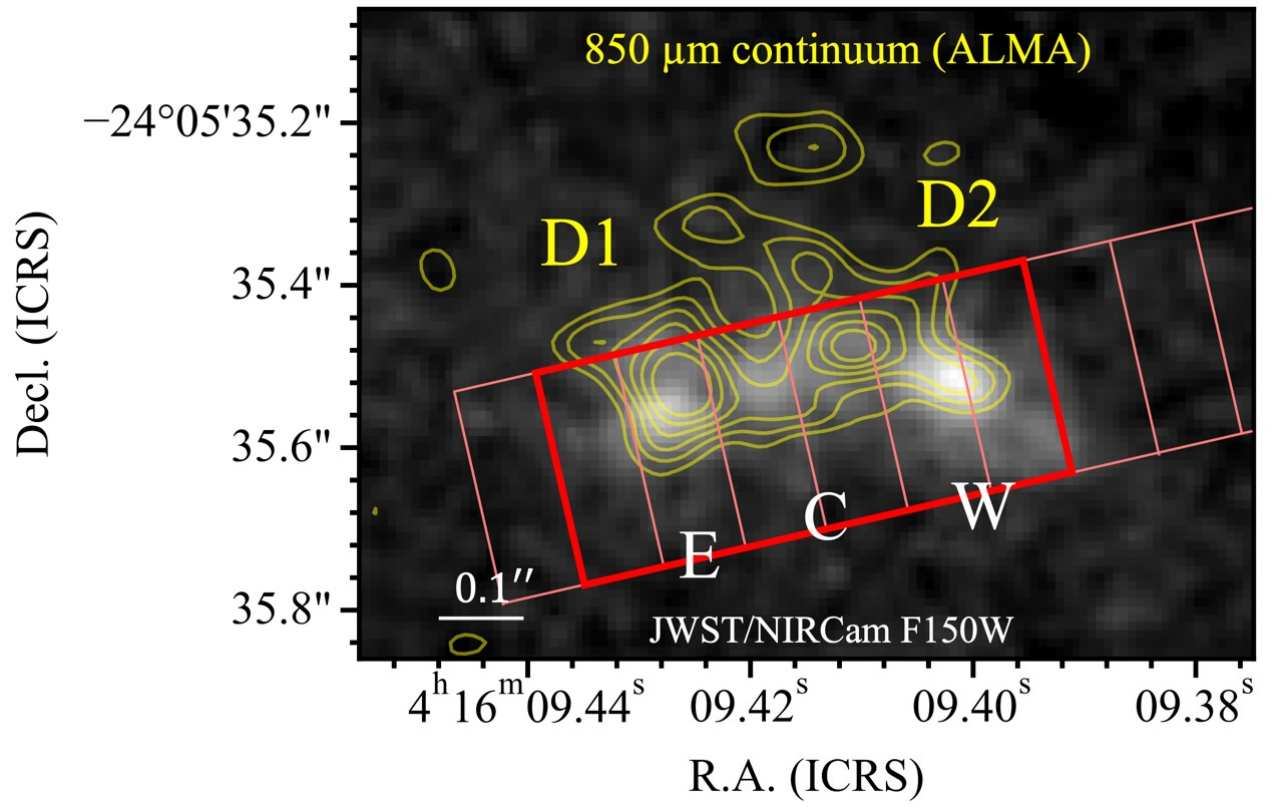}
    \caption{\ckt{\textit{JWST}/NIRSpec F150W image (rest-frame UV continuum) of Y1 in grayscale with 850$\mu$m continuum contours at 2$\sigma$, 3$\sigma$, 4$\sigma$, 5$\sigma$, 6$\sigma$, and \co{7$\sigma$} levels overlaid in yellow. The rest-frame UV peaks, E, C, and W, are labeled in white, while the dust continuum peaks, D1 and D2, are marked in yellow. The red boxes indicate the slit footprint, with each segment corresponding to an individual spaxel.}}
    \label{fig:y1}
\end{figure}

Figure \ref{fig:y1} shows a \textit{JWST}/NIRCam F150W (rest-frame UV) image of Y1 in grayscale, with the 850$\mu$m continuum contours overlaid in yellow. 
We define three distinct peaks in the F150W image as E, C, and W from east to west. The dust continuum exhibits two peaks, D1 (east) and D2 (west). The red boxes indicate the slit \cn{footprint}, with each segment corresponding to an individual spaxel. 
\cn{The slit footprint shown in the figure is based on the \textit{JWST} Astronomer’s Proposal Tool~\footnote{\url{https://www.stsci.edu/scientific-community/software/astronomers-proposal-tool-apt}}, and the spaxel grid is constructed by shifting the slit minor axis in parallel steps of $0\farcs1$ from the reference position.}

\begin{figure}
    \centering
    \includegraphics[width=\linewidth]{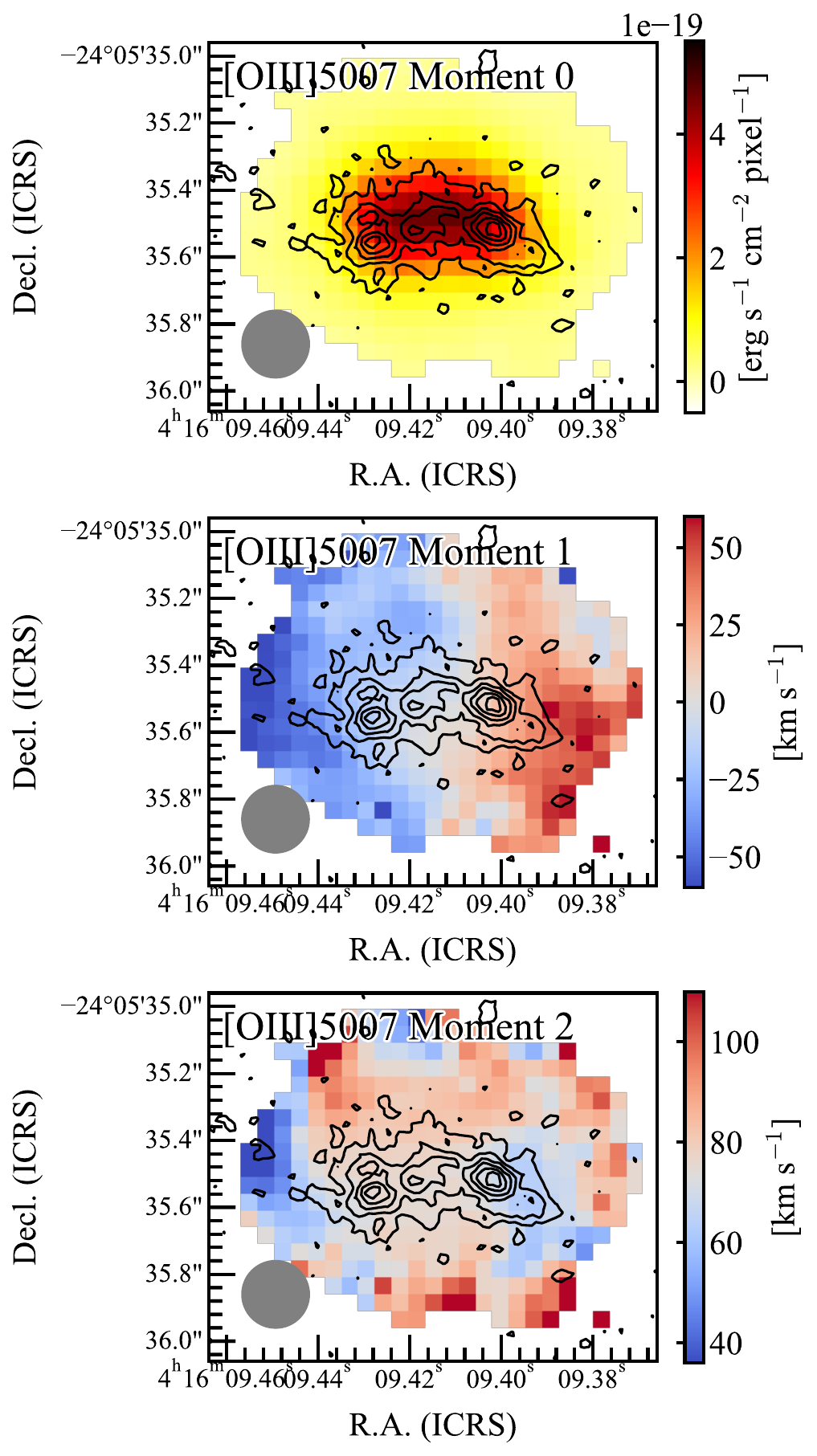}
    \caption{Moment 0 (top), 1 (middle), and 2 (bottom) maps of [\textsc{Oiii}]\,5007. \ca{Pixels with S/N$<5$ are masked.} \ck{The contours show the NIRCam F150W image at 2$\sigma$, 5$\sigma$, 8$\sigma$, 11$\sigma$, 14$\sigma$, and 17$\sigma$ levels. The gray circle indicates the PSF at the observed wavelength of [\textsc{Oiii}]\,5007. The PSF FWHM is estimated using the wavelength-dependent relation derived from serendipitous star in the IFU field of view by \citet{Deugenio2024}. }}
    \label{fig:mommaps}
\end{figure}

Figure \ref{fig:mommaps} presents the moment maps of [\textsc{Oiii}]\,5007, which has the highest signal-to-noise ratio among all detected emission lines. 
\ca{The moment maps are constructed by fitting the emission line profile in each spaxel. The moment 0 map corresponds to the integrated line flux. The moment 1 map represents the line-of-sight velocity, computed from the shift of the line center relative to the systemic velocity defined by the [\textsc{Oiii}]\,88$\mu$m and [\textsc{Cii}]\,158$\mu$m redshift of Y1~\citep[$z=8.312$;][]{Tamura2019,Bakx2020}. The moment 2 map traces the velocity dispersion, derived from the fitted line width after correcting for instrumental broadening.}
The moment 1 map in the central panel shows a velocity gradient \cn{of $\sim\pm60$~km~s$^{-1}$,} with redshifted emission on the eastern side and blueshifted emission on the western side, consistent with rotational motion. A similar velocity gradient is also observed in [\textsc{Cii}]\,158$\mu$m~\citep{Bakx2020}.

\begin{figure}
    \centering
    \includegraphics[width=\linewidth]{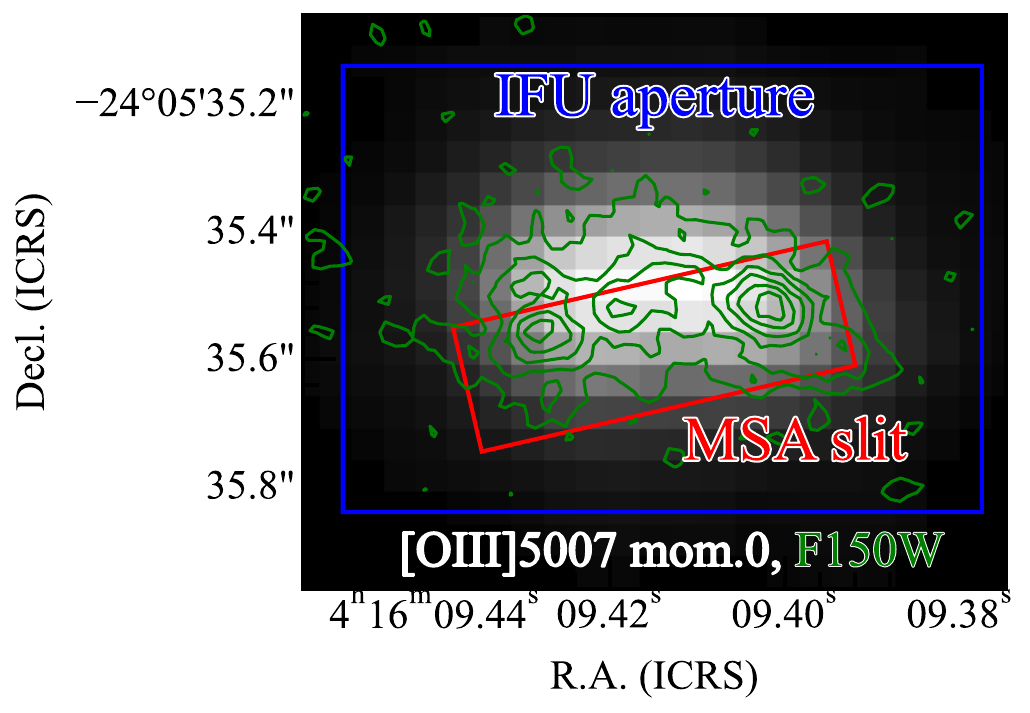}
    \caption{\ckt{Color map of [\textsc{Oiii}]\,5007 moment-0 map, \ck{overlaid with the green contours of NIRCam F150W image at 2$\sigma$, 5$\sigma$, 8$\sigma$, 11$\sigma$, 14$\sigma$, and 17$\sigma$ levels}, and the red (blue) box denotes the MSA slit integration region (IFU aperture) used for spectral extraction. The IFU aperture size is determined from a growth-curve analysis.}}
    \label{fig:aperture}
\end{figure}

\begin{figure}
    \centering
    \includegraphics[width=\linewidth]{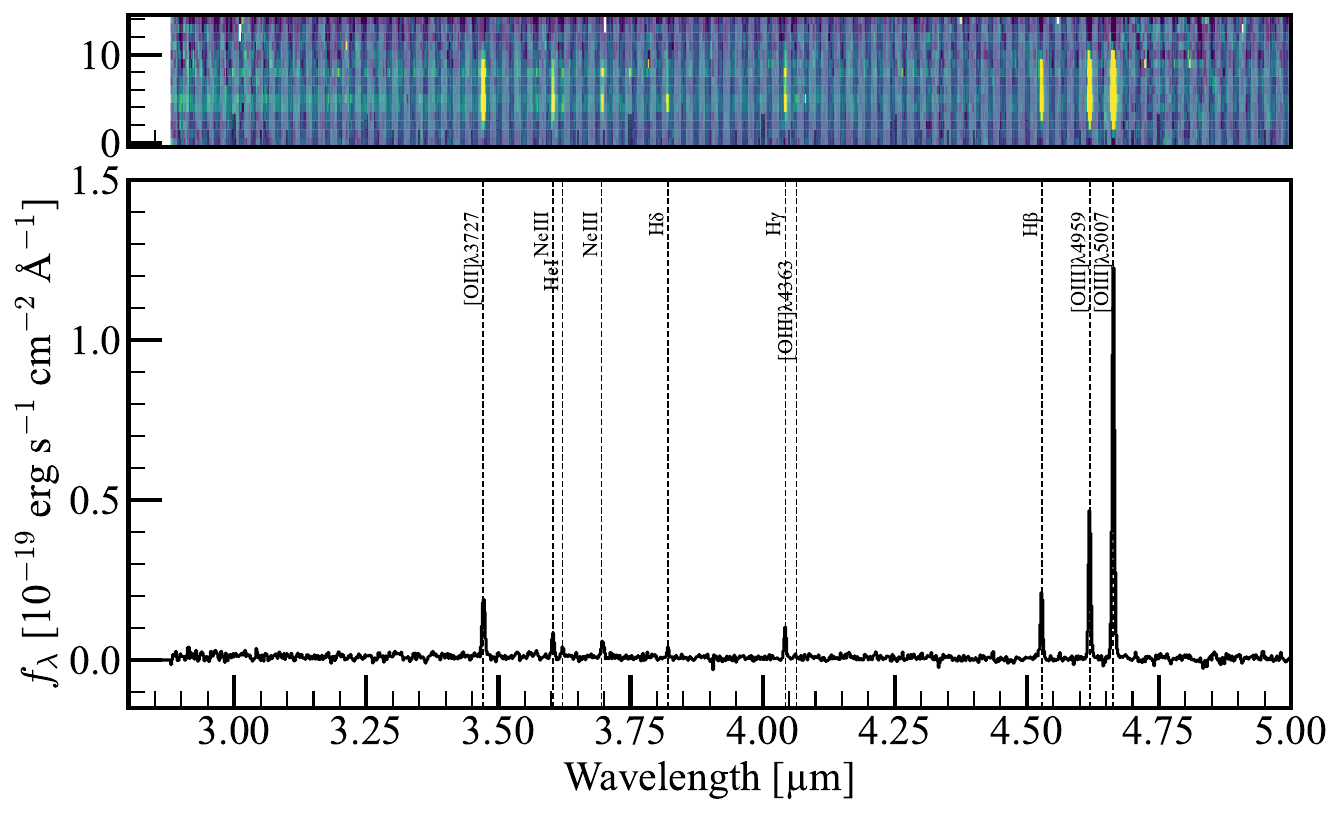}
    \caption{\ckt{Top: \textit{JWST}/NIRSpec MSA two-dimensional spectrum of Y1. Bottom: Extracted one-dimensional spectrum integrated over the spaxels shown by the red box in Figure~\ref{fig:y1}. Emission lines are marked with dashed vertical lines. Flux density is shown in units of $10^{-19}\ \mathrm{erg\ s^{-1}\ cm^{-2}\ \AA^{-1}}$ as a function of observed wavelength.}}
    \label{fig:msa_spec}
\end{figure}

\begin{figure}
    \centering
    \includegraphics[width=\linewidth]{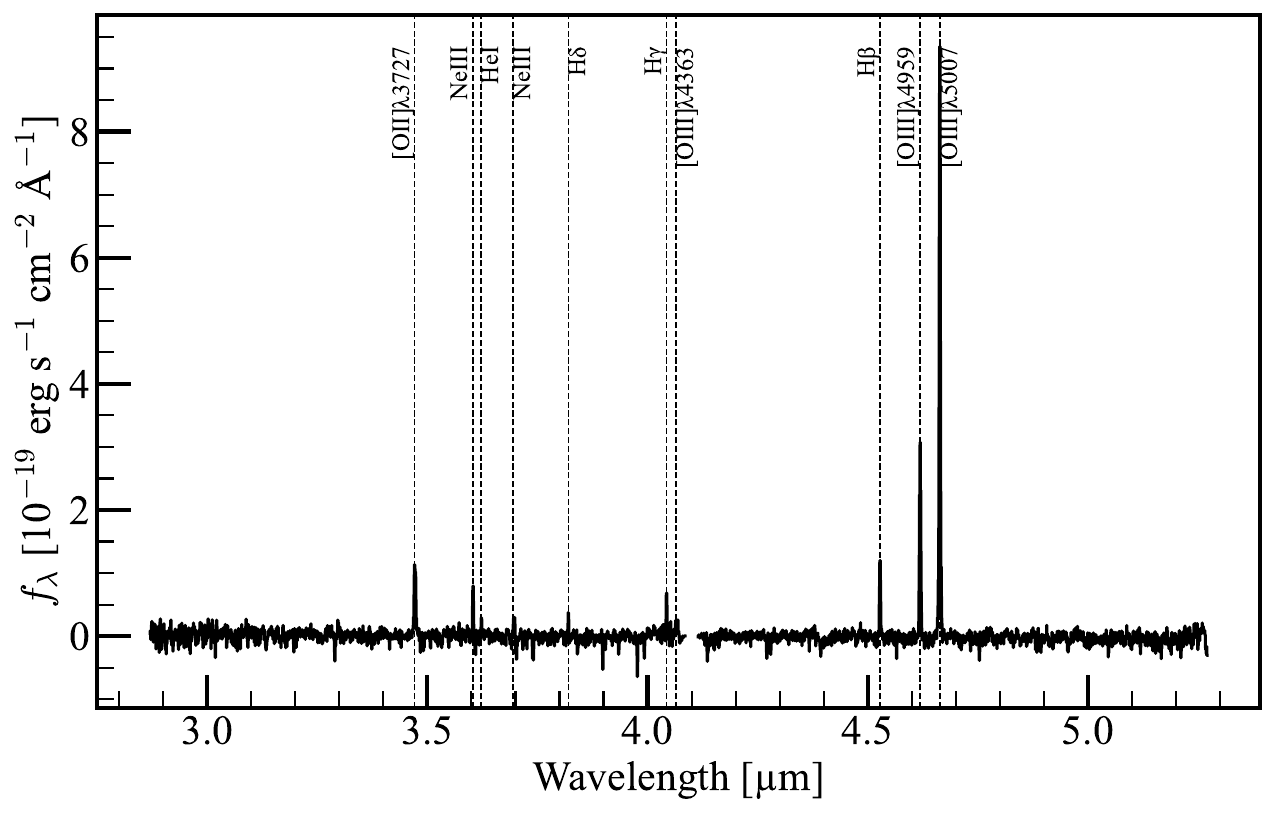}
    \caption{\ckt{\textit{JWST}/NIRSpec IFU one-dimensional spectrum integrated over the blue box in Figure~\ref{fig:aperture}. Emission lines are marked with dashed vertical lines. Flux density is shown in units of $10^{-19}\ \mathrm{erg\ s^{-1}\ cm^{-2}\ \AA^{-1}}$ as a function of observed wavelength.}}
    \label{fig:ifu_spec}
\end{figure}

\cb{The red box in Figure~\ref{fig:aperture} indicates the slit spaxels used to construct the MSA one-dimensional spectrum, corresponding to the thick red box in Figure~\ref{fig:y1}.} The one-dimensional spectrum integrated over these spaxels is shown in the bottom panel of Figure~\ref{fig:msa_spec}. The top panel presents the two-dimensional spectrum, where the vertical axis corresponds to spaxel position, the horizontal axis to wavelength, and the color scale to flux density. Emission lines are marked with dashed vertical lines. The emission-line profiles of H$\beta$ and [\textsc{Oiii}]\,5007 are analyzed in Section~\ref{subsec:broadline}.

\cd{The blue box in Figure~\ref{fig:aperture} indicates the aperture used to extract the IFU-integrated spectrum. The colormap shows the same [\textsc{Oiii}]~$\,5007$ moment 0 map as in Figure~\ref{fig:mommaps}. The IFU aperture is determined from 90\% coverage of [\textsc{Oiii}]~$\,5007$ from the growth curve of the [\textsc{Oiii}]~$\,5007$ flux, with a fixed aspect ratio of 3:4. The resulting aperture size is $\mathrm{R.A.}\times\mathrm{Decl.} = 0\farcs86 \times 0\farcs64$.
\cn{We masked spaxels with flux densities below $-10\sigma$ over the wavelength range $4.0394$--$4.0427~\mu$m in the northeastern region of Y1 to exclude strong negative artifacts, where $\sigma$ is the flux-density uncertainty per spaxel estimated from the line-free channels.} 
The IFU-integrated spectrum extracted within this aperture is shown in Figure~\ref{fig:ifu_spec}. Emission lines are marked with dashed vertical lines. 
The MSA integration region (red box) covers less than 30\% of the total [\textsc{Oiii}]$\,5007$ flux. We therefore use the IFU data to derive the global physical properties of Y1.
}

\subsection{AGN Signatures}
\subsubsection{Broad H$\beta$ Emission}
\label{subsec:broadline}
\begin{figure*}
    \centering
    \includegraphics[width=.7\linewidth]{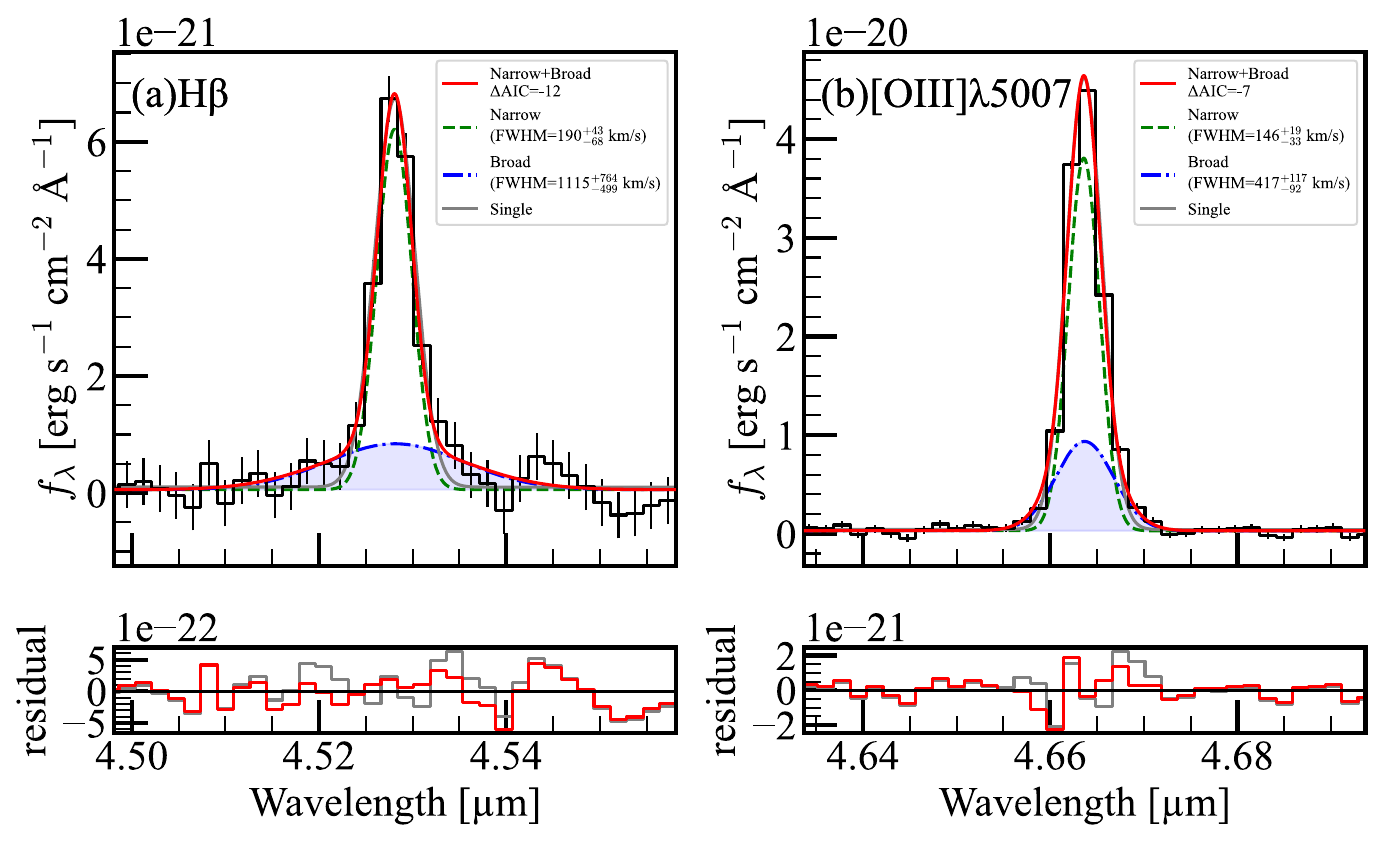}
    \caption{Emission-line profile fitting of (a)H$\beta$ and (b)[\textsc{Oiii}]\,5007, \cj{extracted from the spaxel marked by red box in Figure~\ref{fig:BHpos}}. The black stepped curves and error bars show the observed spectra and their 1$\sigma$ uncertainties. The green dashed, blue dashed-dotted, and red solid curves represent the narrow, broad, and combined narrow$+$broad Gaussian profiles, respectively. The gray curves show the single-Gaussian fits. The bottom panels show the residuals. } 
    
    \label{fig:broadline}
\end{figure*}

\begin{figure}
    \centering
    \includegraphics[width=\linewidth]{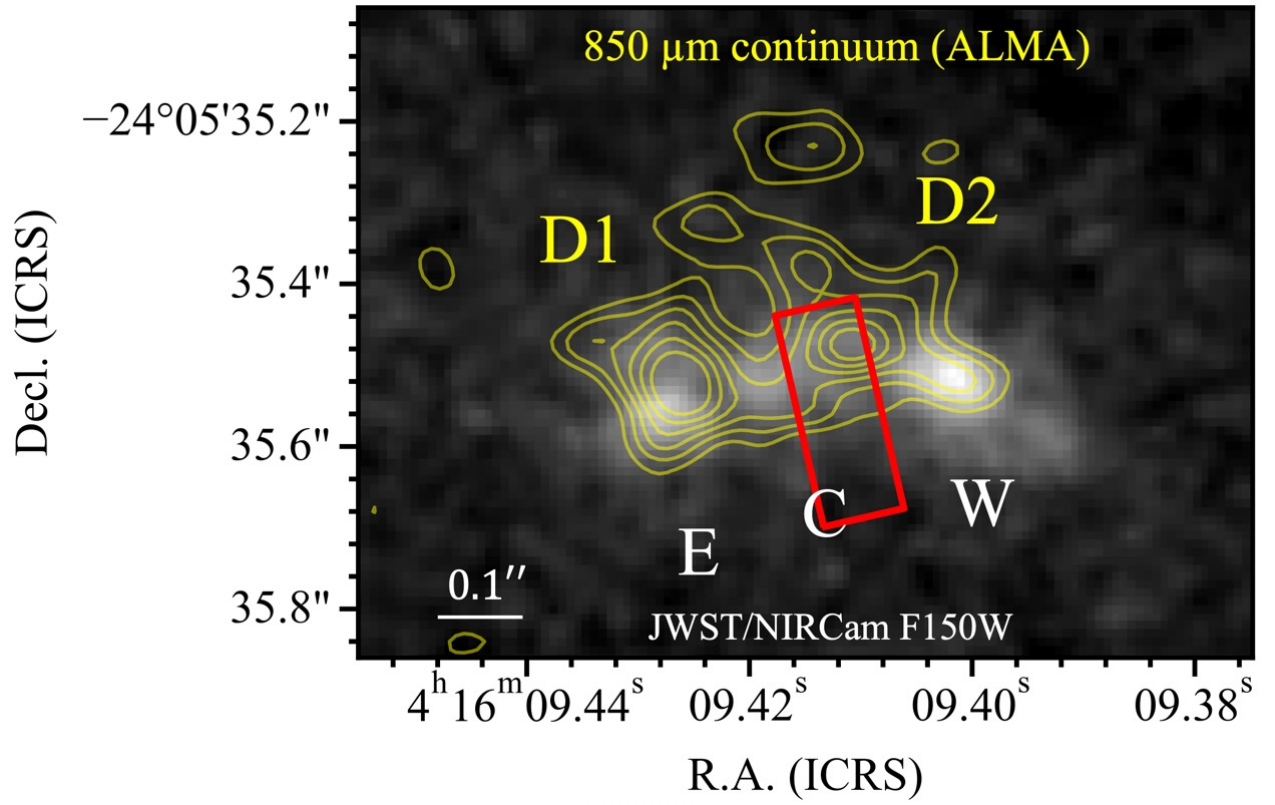}
    \caption{\ckt{\textit{JWST}/NIRCam F150W image of Y1 overlaid with ALMA 850~$\mu$m continuum contours \cc{at 2$\sigma$, 3$\sigma$, 4$\sigma$, 5$\sigma$, 6$\sigma$, \co{and 7$\sigma$} levels} in yellow. The rest-frame UV peaks, E, C, and W, are labeled in white. The dust continuum peaks, D1 and D2, are marked in yellow. The red box indicates the position of the spaxel with broad H$\beta$ emission.}
    \label{fig:BHpos}}
\end{figure}

We perform multi-component Gaussian fitting to search for broad components in the \ck{H$\beta$ line.} % emission lines. 
This analysis is based on \textit{JWST}/NIRSpec MSA data, which provide a higher signal-to-noise ratio than the NIRSpec/IFU observations. 
A broad component is added when the Akaike Information Criterion~\ca{\citep[\cc{AIC;}][]{Akaike1974}} decreases by more than \cb{10}, \cn{$\Delta$AIC$=$AIC$_{\rm double}-$AIC$_{\rm single}<-10$}\ck{, following the statistical model-selection criterion of \citet{Burnham2004}}. 
Figure~\ref{fig:broadline}(a) shows the double-Gaussian fit to the H$\beta$ line for the spaxel corresponding to the region indicated by the red box in Figure~\ref{fig:BHpos}, \cl{which is the only spaxel satisfying the criterion of $\Delta$AIC$<-10$}. The velocity FWHM of the broad H$\beta$ component is $1100^{+800}_{-500}$~km~s$^{-1}$, exceeding the typical threshold used to identify AGN, 1000~km~s$^{-1}$~\citep[e.g.,][]{Harikane2023b,Mathee2024,Kiyota2025}.
Figure~\ref{fig:broadline}(b) shows the double component fit of [\textsc{Oiii}]\,5007 ($\Delta \mathrm{AIC}=-7$) for the same spaxel. The velocity FWHM of the broad [\textsc{Oiii}]\,5007 component is $420^{+120}_{-100}$~km~s$^{-1}$, narrower than that of H$\beta$ \co{at the $\sim1\sigma$ level}. \cf{Broad [\textsc{Oiii}]\,5007 emission is commonly associated with outflows, whereas the line is expected to be suppressed in the high-density broad-line region ($n_e=10^{9\textrm{--}11}$~cm$^{-3}$; \citealt{Osterbrock1986}) because its critical density ($n_{\rm crit}=6.8\times10^5$~cm$^{-3}$;~\citealt{Osterbrock2006}) is far below typical broad-line region densities. If the broad H$\beta$ emission originated from outflowing gas, a comparable velocity width would be expected in [\textsc{Oiii}]\,5007. The observed difference in line width therefore argues against an outflow origin for the broad H$\beta$ component.} 
\cn{In addition, \citet{Xu2025} found that galaxies without AGN typically exhibit outflow velocities of at most $\sim500$~km~s$^{-1}$, further arguing against an origin of the broad H$\beta$ component solely from outflows.} 
\cb{Instead, this suggests that the broad H$\beta$ line is emitted from a compact and dense broad-line region in the vicinity of an AGN.} \cd{The \cn{observed} H$\beta$ broad-line flux is $f_{\rm H\beta,broad} = 1.43^{+0.62}_{-0.60}\times10^{-19}~\mathrm{erg~s^{-1}~cm^{-2}}$}. \ca{\cn{After correcting for lensing magnification,} the black hole mass is estimated to be \cn{$M_{\rm BH} = 1.6^{+3.0}_{-1.0} \times 10^6~M_\odot$} using the relation in \citet{Greene2005}.} 
\co{Given the stellar mass of $M_{\star}\sim10^9 M_\odot$~\citep{Harshan2024}, the inferred black hole mass of $M_{\rm BH}\sim10^6 M_\odot$ corresponds to $M_{\rm BH}/M_{\star}\sim10^{-3}$, consistent with the $M_{\rm BH}$--$M_{\star}$ scaling relations in local galaxies~\citep[e.g.,][]{Greene2016}, suggesting that the black hole is neither significantly overmassive nor undermassive.} 
%\cg{We assume that the emission is not significantly affected by dust attenuation based on the Balmer line ratio, ${\rm H}\gamma/{\rm H}\beta = 0.47^{+0.05}_{-0.04}$, and therefore no dust attenuation correction is applied.} 
%\ck{The fluxes used to derive this ratio are measured after masking spaxels where H$\gamma$ falls within the wavelength gap.}
\ca{Figure~\ref{fig:BHpos} shows that the red box, corresponding to the spaxel where broad-line component is detected, coincides with the western dust continuum peak, D2. 
The relation between the elevated dust temperature and AGN activity is discussed in Section~\ref{sec:dustheating}.}

\begin{figure}[h]
    \centering
    \includegraphics[width=.8\linewidth]{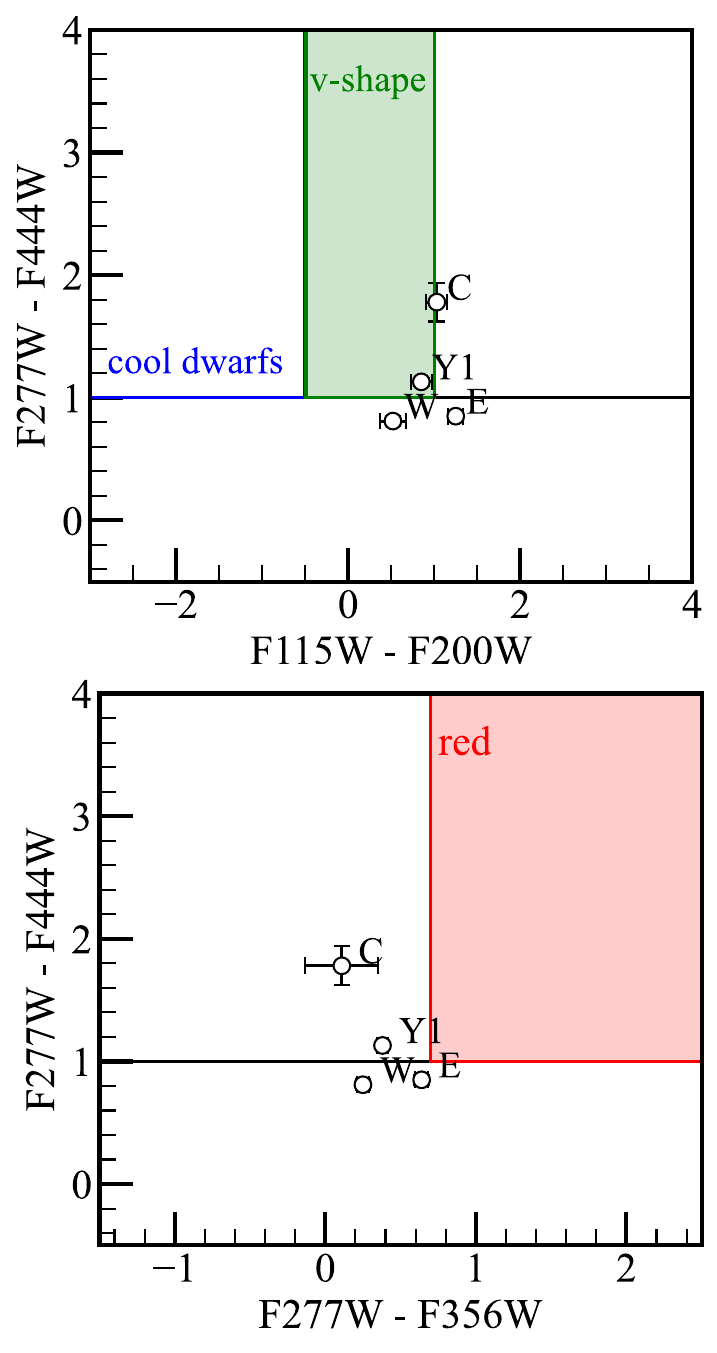}
    \caption{Top: Color criterion used to identify the characteristic V-shaped spectral energy distributions (SEDs) of LRDs in the F277W $-$ F444W versus F115W $-$ F200W plane. The green shaded region indicates the selection window for V-shaped spectral sources, while the blue region marks the locations of cool dwarfs. Bottom: Color criterion used to identify red sources in the F277W $-$ F444W versus F277W $-$ F356W plane. The red shaded region denotes the color-selection criterion for red sources. \cd{LRD candidates are required to satisfy the V-shaped, red, and compactness criteria. The labeled points with error bars (Y1, C, W, and E) show the measured colors of Y1 and its three clumps.} \cg{Photometric measurements are taken from \citet{Ma2024}. }
    \label{fig:LRDtest}}
\end{figure}

\subsubsection{LRD Test}
\cp{We apply the Little Red Dot (LRD) color selection criteria from \citet{Greene2024} to the NIRCam photometry to test whether Y1 satisfies the LRD selection criteria. The upper panel of Figure~\ref{fig:LRDtest} shows the color criterion used to identify the characteristic V-shaped spectral energy distribution (SED) of LRDs.} 
The \ca{V}-shaped criterion is defined by
\begin{equation}
    \begin{split}
        (-0.5<\mathrm{F115W-F200W}<1.0) \land \\
        (\mathrm{F277W-F444W}>1.0).
    \end{split}
\end{equation}
The lower panel of Figure~\ref{fig:LRDtest} shows the color criterion used to identify red sources,
\begin{equation}
\mathrm{F277W-F356W}>1.0.
\end{equation}
\cp{LRD candidates are required to satisfy both the V-shaped and red color criteria (as well as the compactness criterion)~\citet{Greene2024}. 
No individual clumps or the integrated photometry of Y1 satisfy all of these criteria, suggesting that Y1 is not classified as an LRD.} 

\subsubsection{MEx Diagram}
\begin{figure}
    \centering
    \includegraphics[width=.9\linewidth]{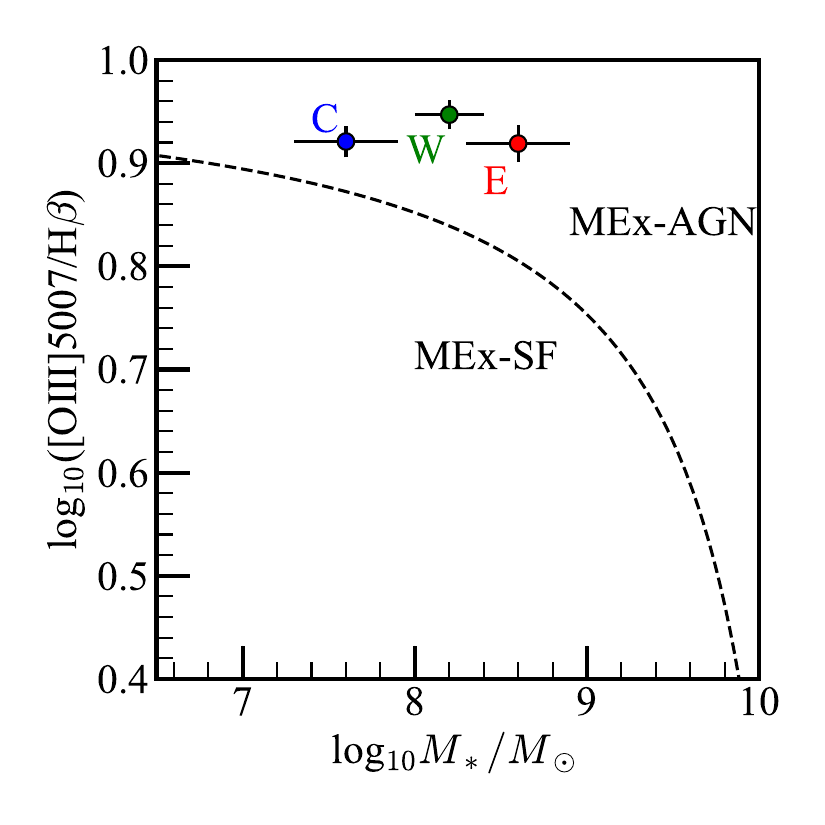}
    \caption{Mass--Excitation (MEx) diagram showing $\log_{10}$([\textsc{Oiii}]\cc{5007}/H$\beta$) versus stellar mass. The dashed curve marks the empirical division between star-forming galaxies (MEx-SF) and AGN-dominated systems (MEx-AGN)~\citep{Juneau2011}. The measured values for clumps E (red), C (blue), and W (green) are shown with uncertainties. \cn{Stellar masses are taken from \citet{Harshan2024}.}}
    \label{fig:mex}
\end{figure}

\ckt{In Section~\ref{subsec:broadline}, we find a broad H$\beta$ line component at the position corresponding to D2, whereas no significant broad-line component is detected at the position of D1. However, the absence of a broad-line component does not rule out the presence of an AGN. }
\cp{To investigate the ionization properties of all clumps including those without detected broad-line components, we examine their locations on the Mass-Excitation (MEx) diagram, which empirically separates AGN from star-forming galaxies using the [\textsc{Oiii}]\,5007/H$\beta$ ratio and stellar mass~\cc{\citep{Juneau2011}}.} \ck{For $\log_{10}(M_*/M_\odot) < 9.9$, the dividing curve is given by $\log_{10}([\textsc{Oiii}]\,5007/\mathrm{H}\beta) = 0.37/(\log_{10}(M_*/M_\odot)-10.5)+1$ \citep{Juneau2011}.} Figure \ref{fig:mex} shows the location of each clump on this plane. The line ratios are derived from the NIRSpec IFU spectra extracted with a 0$\farcs$08 aperture centered on each clump. For stellar masses, we adopt the SED-fitting results from \citet{Harshan2024}. All three clumps lie in the MEx-AGN regime, suggesting high ionization conditions consistent with AGN activity. However, we note that this diagnostic was calibrated using galaxies at $0.3 < z < 1$ and, similar to the BPT diagram, may not be directly applicable to the ISM conditions of high-redshift galaxies~\cj{\citep[e.g.,][]{Kewley2013,Steidel2014,Shapley2015,Sanders2016,Topping2020}. }

\subsection{Metallicity}

\subsubsection{\ch{Direct Method}}
\label{sec:zte}
\begin{figure}
    \centering
    \includegraphics[width=.7\linewidth]{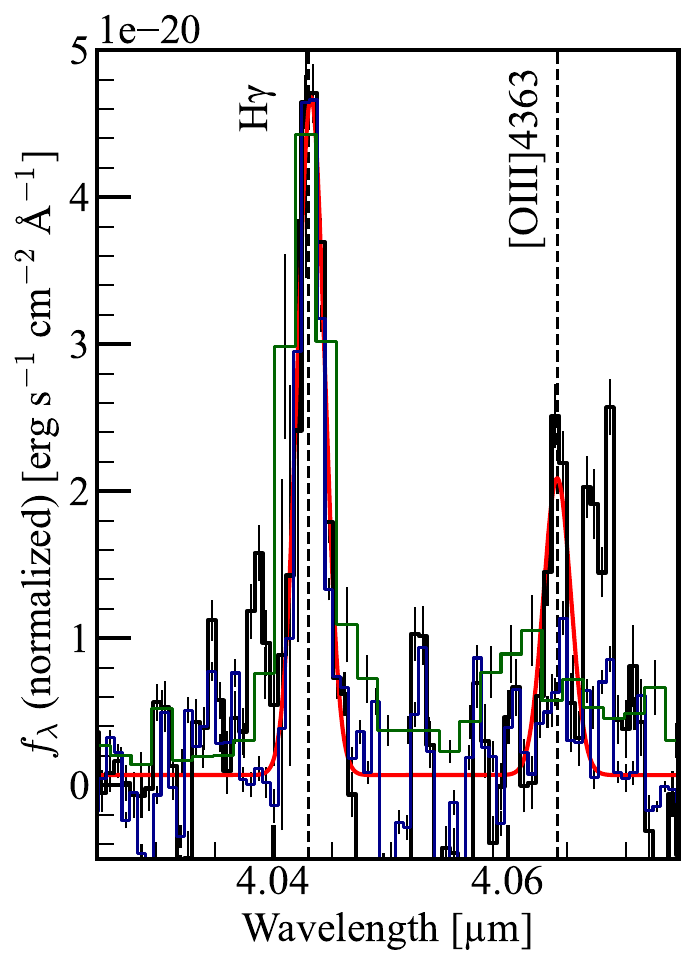}
    \caption{Spectra and best-fit profiles of H$\gamma$ and [\textsc{Oiii}]\,4363. The IFU-integrated spectrum, MSA spectrum, and the IFU spectrum integrated over the MSA aperture are shown in \cb{thick} black, green, and blue, respectively. \ce{The MSA spectrum and the IFU spectrum integrated over the MSA aperture are normalized to the IFU-integrated spectrum at the peak of H$\gamma$.} The red curve shows the best-fit model to the IFU-integrated spectrum. \cn{The broader line widths in the MSA spectrum compared to the IFU spectra are due to the larger LSF FWHM in the medium-grating MSA mode relative to the high-resolution IFU observations.}}
    \label{fig:4363}
\end{figure}

We derive the gas-phase metallicity using the \ch{direct method}. Figure~\ref{fig:4363} presents the spectra of H$\gamma$ and [\textsc{Oiii}]\,4363. The IFU-integrated spectrum, the MSA spectrum, and the IFU spectrum integrated over the MSA aperture are shown as thick black, green, and blue stepped curves, respectively. \cd{The MSA spectrum and the IFU spectrum integrated over the MSA aperture are normalized to the IFU-integrated spectrum at the peak of H$\gamma$.}
\cd{The broader line widths in the MSA spectrum compared to the IFU spectrum arise from the larger LSF FWHM of the medium-grating MSA mode relative to the higher-resolution IFU observations.}
To characterize the global properties of the galaxy, we derive the metallicity from the IFU-integrated spectrum. \cg{For the IFU-integrated spectrum, southeastern spaxels without [\textsc{Oiii}]\,4363 coverage due to wavelength gaps are masked prior to integration \cn{for this analysis}. The resulting flux loss corresponds to 27\% of the total [\textsc{Oiii}]\,5007 flux.}
The red curve in the Figure~\ref{fig:4363} represents the best-fit profile to the IFU-integrated spectrum. \cg{The velocity width and redshift of [\textsc{Oiii}]\,4363 are fixed to those of [\textsc{Oiii}]\,5007.}
We derive an electron temperature for the [\textsc{Oiii}] region from the [\textsc{Oiii}] line ratio, obtaining $T_e^{[\textsc{Oiii}]}=17300\pm1500$~K. \ca{Using the relation between $T_e^{[\textsc{Oiii}]}$ and the electron temperature of the [\textsc{Oii}] region \citep[$T_e^{[\textsc{Oii}]}=0.7\times T_e^{[\textsc{Oiii}]}+3000$~K;][]{Stasinska1982,Campbell1986}, we estimate $T_e^{[\textsc{Oii}]}=15100^{+1100}_{-1000}$~K. Using these electron temperatures and the fluxes of [\textsc{Oii}] and [\textsc{Oiii}],} we derive a gas-phase metallicity of $12+\log{(\rm O/H)}=7.86^{+0.09}_{-0.08}\sim0.15~Z_\odot$. The calculation is performed using \texttt{PyNeb}~\citep{Luridiana2015}, \ca{assuming an electron density of $n_e = 100~\mathrm{cm^{-3}}$}. \ca{We neglect ions with ionization states higher than $\mathrm{O^{2+}}$.}

\subsubsection{Strong Line Method}
\begin{figure}
    \centering
    \includegraphics[width=\linewidth]{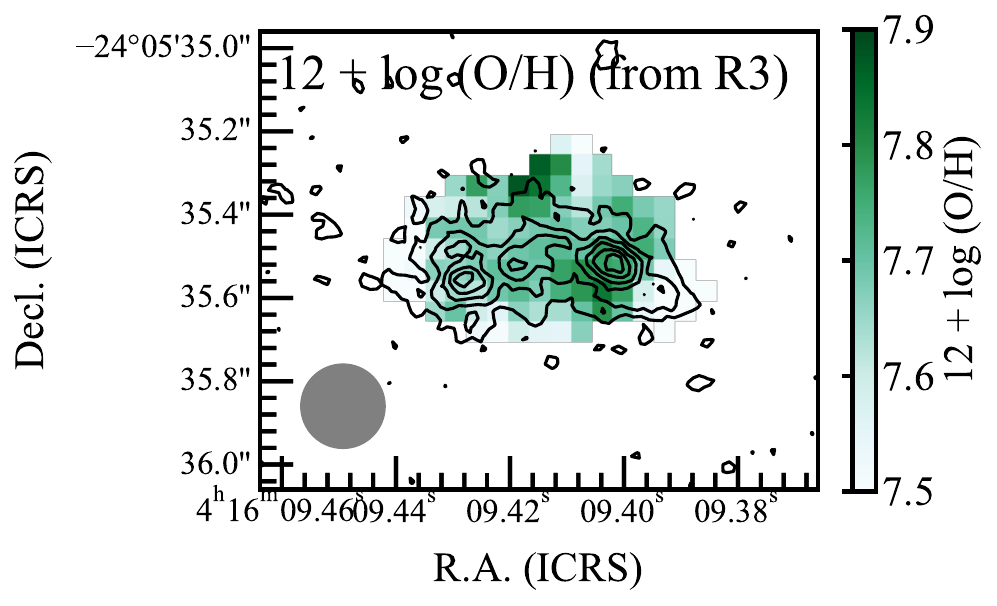}
    \caption{\cj{Metallicity $12 + \log(\mathrm{O/H})$ derived from $R3$ using the \citet{Hirschmann2023} relation. Spaxels with S/N$<5$ are masked. The contours correspond to the NIRCam F150W image at 2$\sigma$, 5$\sigma$, 8$\sigma$, 11$\sigma$, 14$\sigma$, and 17$\sigma$ levels. The gray circle indicates the PSF at the observed wavelength of [\textsc{Oiii}]\,5007. Because the PSF FWHMs at the observed wavelengths of [\textsc{Oiii}]\,5007 and H$\beta$ differ by only $\sim0\farcs01$, no PSF matching is applied.}}
    \label{fig:R3logOH}
\end{figure}

We \ca{also} estimate the gas-phase metallicity using \cb{a} strong line method. % \ca{to investigate spatial distribution of metallicity}.
Based on fluxes measured from the \textit{JWST}/NIRSpec IFU data, we compute the $R3$ index (=log$_{10}$([\textsc{Oiii}]\,5007/H$\beta$)). Metallicity is then inferred from $R3$ using calibrations based on empirical relations~\citep{Nakajima2022} and cosmological simulations~\citep{Hirschmann2023}. From the IFU-integrated spectrum, we obtain $R3 = 0.90 \pm 0.02$, \ca{where the uncertainty is estimated by propagating the posterior samples of the [\textsc{Oiii}] and H$\beta$ fluxes.} Using the \citet{Hirschmann2023} relation, this $R3$ value corresponds to a metallicity of $12+\log(\mathrm{O/H})=7.73\pm0.04$, \cj{with the spatial distribution shown in Figure \ref{fig:R3logOH}.} The \citet{Nakajima2022} calibration does not intersect with this $R3$ value (lower than the observation), but its maximum corresponds to a metallicity of \ce{$12+\log(\mathrm{O/H})=7.94$}. From these results, we estimate a metallicity of $12 + \log(\mathrm{O/H}) \approx 7.7$--$8.0$ based on the $R3$ index. This is consistent with the results from the direct method within uncertainties.

\begin{figure*}
    \centering
    \includegraphics[width=\linewidth]{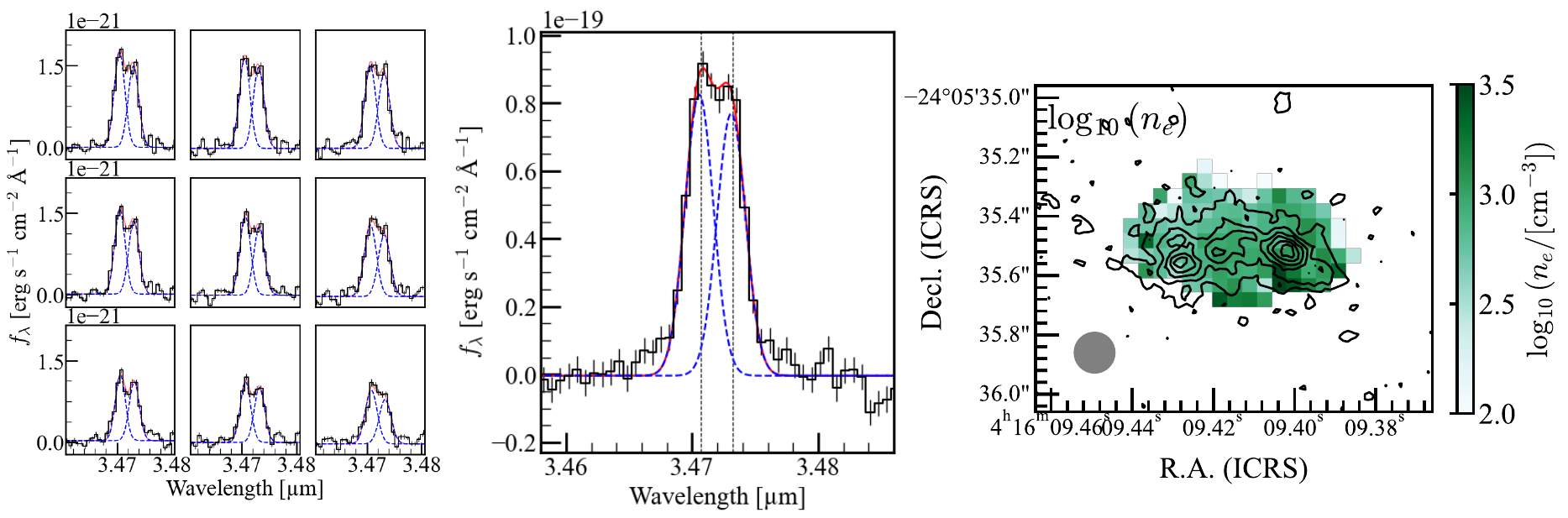}
    \caption{
    Left: Example spaxel-wise Gaussian fits to the [\textsc{Oii}]\,3726,3729 doublet for a $3\times3$ grid. Center: Integrated [\textsc{Oii}] spectrum constructed by summing only spaxels with $\mathrm{S/N} > 5$. Solid red curves show the best-fit Gaussian profile to the [\textsc{Oii}] doublet, and blue dashed curves show the individual components. \cj{Right: Spatial distribution of electron density on a logarithmic scale. Spaxels with $\mathrm{S/N} < 5$ are masked. The contours correspond to the NIRCam F150W image at 2$\sigma$, 5$\sigma$, 8$\sigma$, 11$\sigma$, 14$\sigma$, and 17$\sigma$ levels.} \cl{The gray circle indicates the PSF at the observed wavelength of [\textsc{Oii}] doublet. }}
    \label{fig:ne_o2}
\end{figure*}

\subsection{Electron Density}

\subsubsection{[\textsc{Oii}]\,3729/3726}

We derive the electron density from the [\textsc{Oii}] doublet ratio using high spectral resolution \textit{JWST}/NIRSpec IFU data. The electron density is computed using \texttt{PyNeb}. We adopt the electron temperature derived in Section~\ref{sec:zte}.
The left panel of Figure~\ref{fig:ne_o2} shows examples of spaxel-wise [\textsc{Oii}] doublet spectra and the corresponding fitting results. The two components of the doublet are clearly resolved. \ca{We perform fitting for each spaxel and construct a high-S/N integrated spectrum by summing only spaxels with [\textsc{Oii}] detections at \ck{S/N$>5$}}. The spectrum and its best-fit model are shown in the \ca{center panel of} Figure~\ref{fig:ne_o2}. From this spectrum, we obtain an electron density of \ck{$n_e = 730^{+150}_{-140}~\mathrm{cm^{-3}}$}. \ck{
The right panel of Figure~\ref{fig:ne_o2} shows the spatial distribution of the electron density on a logarithmic scale for the spaxels used to construct the high-S/N integrated spectrum. The entire system exhibits relatively high electron densities of several hundred to several thousand cm$^{-3}$.} \cd{We also derive the electron density from the IFU-integrated spectrum, and obtain $n_e = 862^{+310}_{-242}~\mathrm{cm^{-3}}$, consistent within uncertainties with the values derived from high-S/N spectrum.} 

\subsubsection{[\textsc{Oiii}]\,88$\mu$m/5007}
\label{sec:o3rat}

\begin{figure*}
    \centering
    \includegraphics[width=.9\linewidth]{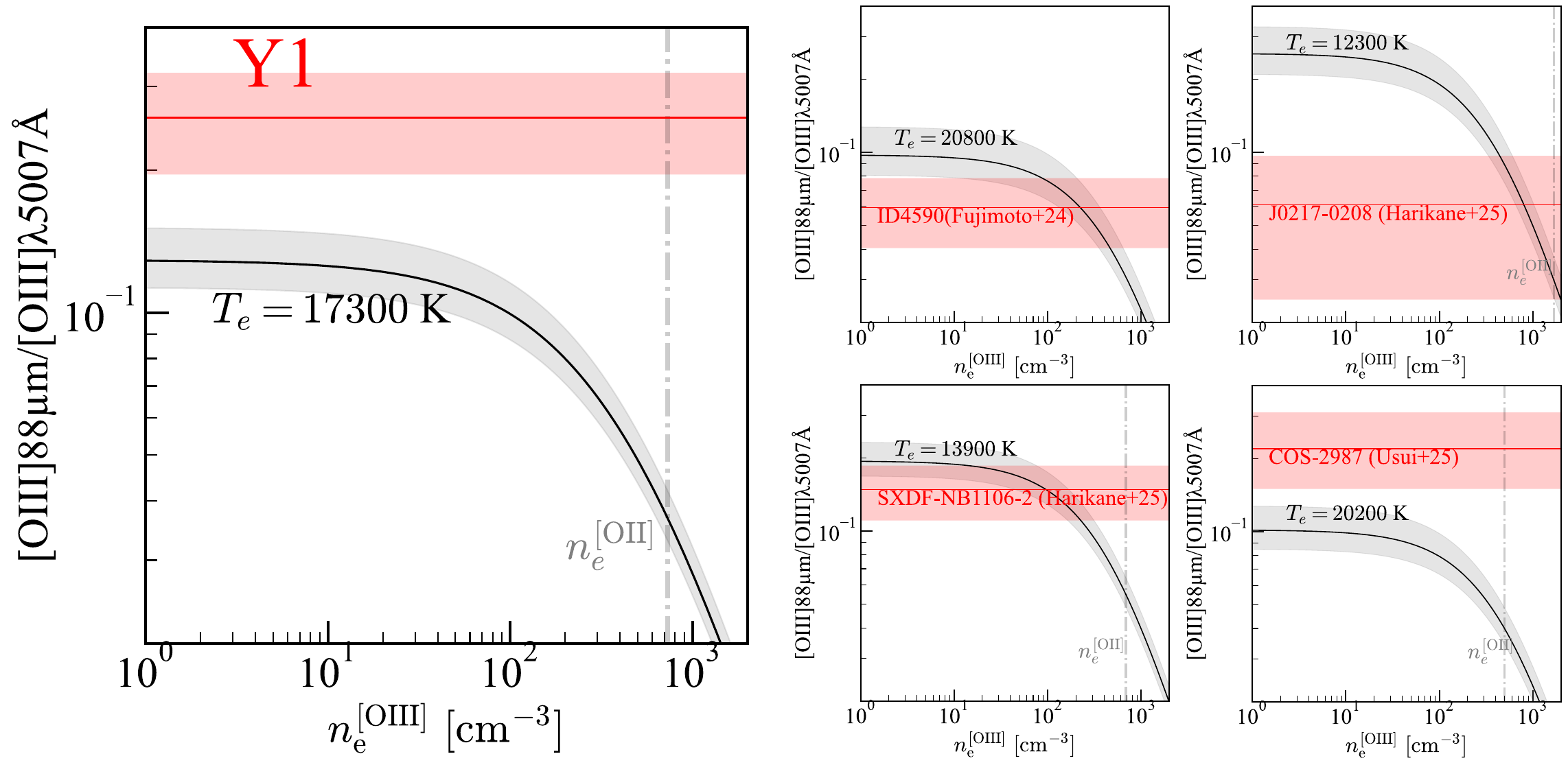}
    \caption{Left: Ratio of [\textsc{Oiii}]\,88$\mu$m to [\textsc{Oiii}]\,5007 as a function of electron density ($n_e^{\rm [OIII]}$). The black curve shows the theoretical relation assuming an electron temperature of $T_e = 17300\pm1500$~K, corresponding to the value measured for Y1, and the gray shaded region represents the uncertainty. The red horizontal line and shaded region indicate the observed ratio and its uncertainty for Y1. The vertical dashed line marks the electron density inferred from the optical [\textsc{Oii}] diagnostic. \cq{Right: Same as the left panel, but for ID4590~\citep[top left;][]{Fujimoto2024}, J0217-0208~\citep[top right;][]{Harikane2025}, SXDF-NB1106-2~\citep[bottom left;][]{Harikane2025}, and COS-2987~\citep[bottom right;][]{Usui2025}.}}
    \label{fig:o3rat_ne}
\end{figure*}

We derive the rest-frame FIR to optical line ratio [\textsc{Oiii}]\,88$\mu$m/[\textsc{Oiii}]\,5007 by combining ALMA and IFU observations. For the [\textsc{Oiii}]\,88$\mu$m flux, we use the value reported by~\citet{Tamura2019}, $0.66\pm0.16$~Jy~km~s$^{-1}$. \ca{The [\textsc{Oiii}]\,5007 flux is measured from the IFU-integrated spectrum, $(3.10\pm0.02)\times10^{-17}~\mathrm{erg~s^{-1}~cm^{-2}}$.} From these values, we obtain an [\textsc{Oiii}] line ratio of [\textsc{Oiii}]\,88$\mu$m/[\textsc{Oiii}]\,$5007=0.26\pm0.06$. This value and uncertainty are shown in the left panel of Figure~\ref{fig:o3rat_ne}. The theoretical relation between the [\textsc{Oiii}] line ratio and electron density at $T_{e}=17300$~K (the electron temperature derived \cc{using \ch{direct method}, see Section~\ref{sec:zte}}) is also plotted \ca{as the black curve}. The gray shaded region presents the uncertainty propagated from the electron temperature. The figure shows that, under the assumption of a single homogeneous ionized nebula, the observed ratio cannot be reproduced for any electron density. Further discussion of this discrepancy and the right panels of Figure~\ref{fig:o3rat_ne} is given in Section~\ref{sec:o3ratdisc}.
%A detailed discussion of this discrepancy, along with the right panels of Figure~\ref{fig:o3rat_ne}, is presented in Section~\ref{sec:o3ratdisc}. 

\section{Discussion}
\label{sec:5}
\subsection{Sources of Dust Heating}
\label{sec:dustheating}

\begin{figure}
    \centering
    \includegraphics[width=\linewidth]{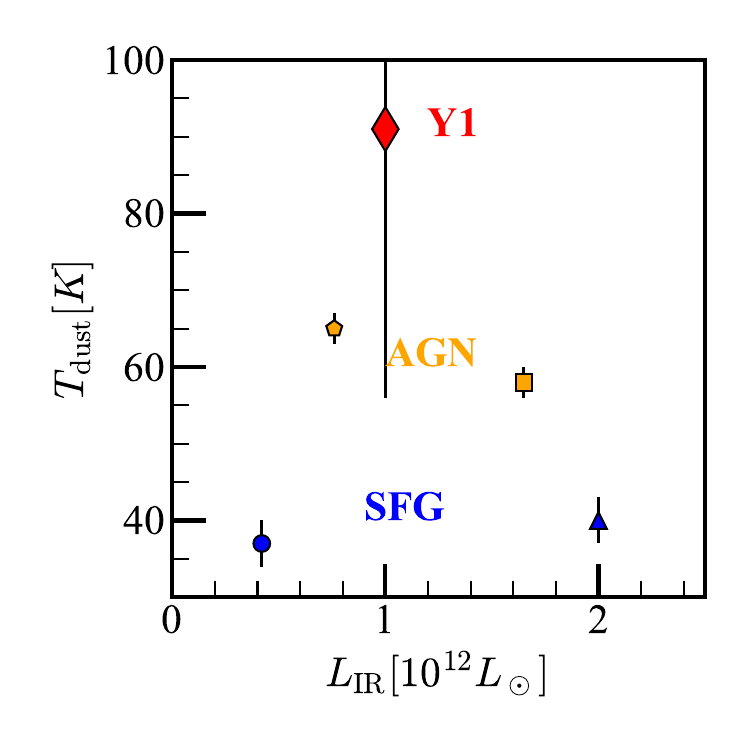}
    \caption{Dust temperature $T_{\rm dust}$ as a function of infrared luminosity ($L_{\rm IR}$). \cd{The red diamond shows Y1 from \citet{Bakx2025}}, compared with \cj{stacked measurements of} $z\sim1$ (blue circle) and $z\sim2$ (blue triangle) star-forming galaxies, silicate AGN (orange square), and featureless AGN (orange pentagon) from~\citet{Kirkpatrick2012}.}
    \label{fig:Td_LIR}
\end{figure}

\begin{figure}
    \centering
    \includegraphics[width=\linewidth]{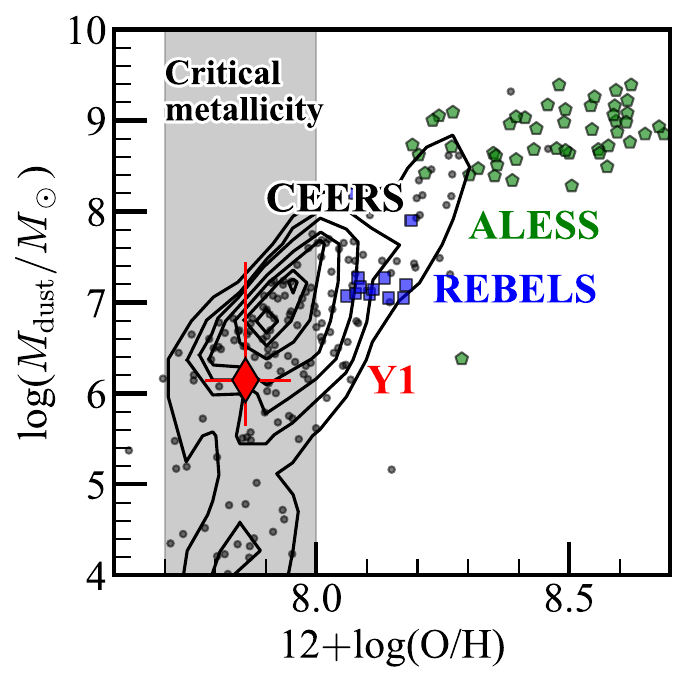}
    \caption{Dust mass as a function of gas-phase metallicity. The red diamond marks Y1. The gray shaded region indicates the \cg{critical metallicity range ($0.1\textrm{--}0.2~Z_\odot$)}. The blue squares and green pentagons represent the \cd{REBELS~\citep[$z\sim$7;][]{Algera2025} and ALESS~\citep[$z\sim$1--6;][]{daCunha2015} samples}, respectively. The black contours and points show the distribution of galaxies \cc{at $4 < z < 11$} from CEERS~\citep[][]{Burgarella2025}. \cn{For all galaxies except for Y1, the metallicities are estimated from stellar masses using the mass--metallicity relation at $z=4$--10~\citep{Nakajima2023}.}
    }
    \label{fig:z_crit}
\end{figure}

\begin{figure*}
    \centering
    \includegraphics[width=1\linewidth]{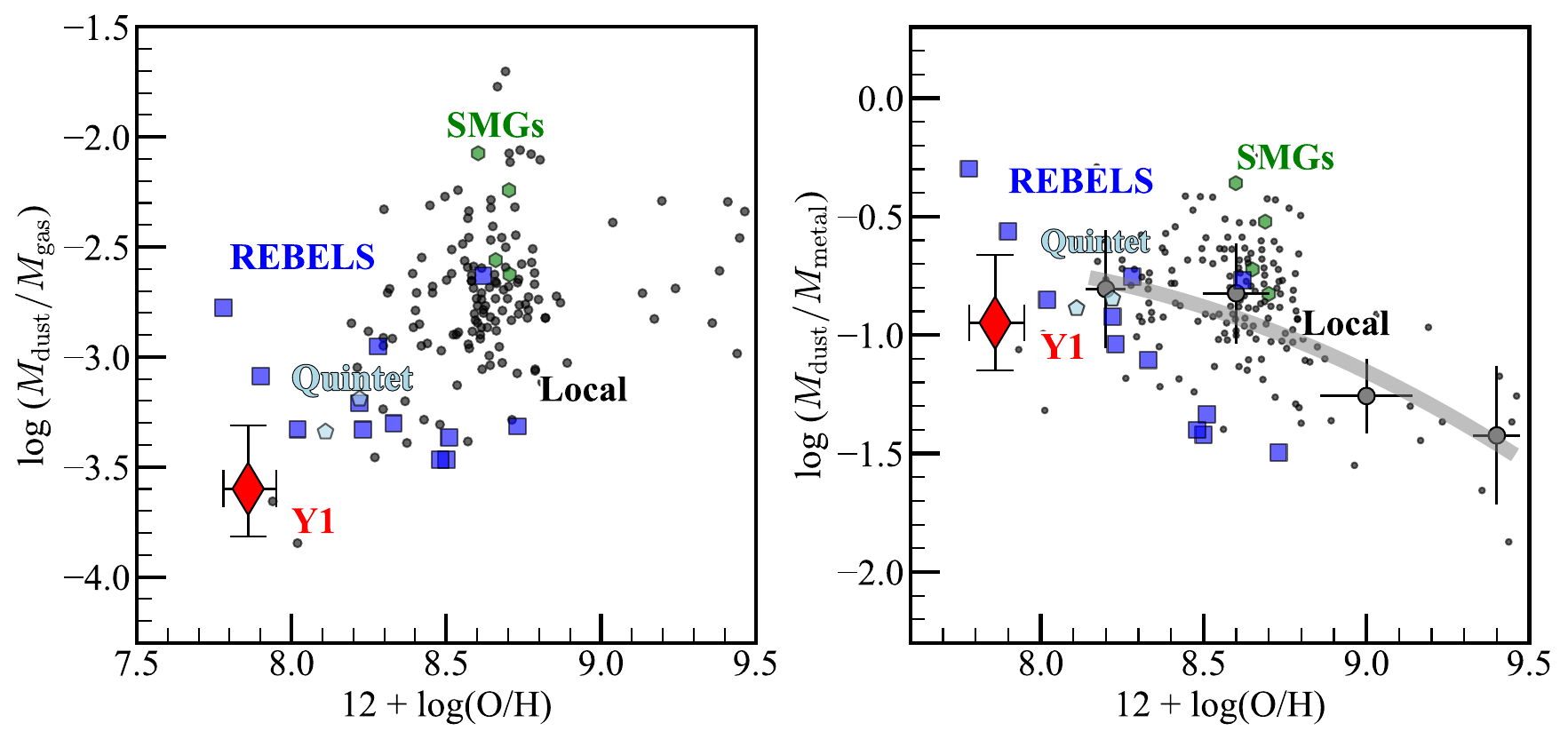}
    \caption{Left: Dust-to-gas ratio, $\log(M_{\rm dust}/M_{\rm gas})$, as a function of gas-phase metallicity, $12+\log(\mathrm{O/H})$. The blue squares, light blue pentagons, green heptagons, and black points show \cd{REBELS galaxies~\citep[$z\sim$7;][]{Algera2025}}, Quintet YD1 and YD4YD6~\citep[$z\sim7.9$;][]{Umehata2025}, submillimeter galaxies (SMGs)~\citep[$z\sim$2;][]{Shapley2020} and local galaxies~\citep{DeVis2019,Casasola2020}, respectively. The red diamond marks Y1. Right: Same as the left panel, but for the dust-to-metal ratio, $\log(M_{\rm dust}/M_{\rm metal})$, on the vertical axis. The gray points indicate the median values of local galaxies at $12+\log(\mathrm{O/H})>8.0$, binned with a width of $\Delta(12+\log(\mathrm{O/H}))=0.4$. The gray curve shows a polynomial fit to these median values.}
    \label{fig:Md_comp}
\end{figure*}

Y1 has been observed with ALMA Bands 3--9, with continuum emission detection in Bands 5--9~\citep{Tamura2019,Tamura2023,Bakx2020,Bakx2025,Jones2024,Harshan2024}. \ca{\citet{Bakx2025} report dust temperature and dust mass of 91$^{+62}_{-35}$~K and 1.4$^{+1.3}_{-0.5}\times10^6~M_\odot$, respectively, based on SED fitting of these observations.} This temperature is higher than typical values inferred for galaxies at $z\gtrsim7$~\citep[40--60~K;][]{Sommovigo2022,Mitsuhashi2024}.

\cd{The origin of such elevated dust temperatures remains under debate. 
A redshift evolution of dust temperature has been reported, with $T_{\rm dust}$ increasing toward higher redshift~\citep{Bethermin2015,Sommovigo2022,Mitsuhashi2024}.
Extrapolation of empirical relations derived at $z \lesssim 7$ predicts dust temperatures of $\sim 40$–$60~\mathrm{K}$ at $z \sim 8.3$. }
\cd{This trend has been interpreted as a consequence of the lower metallicity and dust mass at high-$z$, which reduce dust shielding and enable more efficient absorption of UV radiation~\citep[e.g.,][]{Sommovigo2022}. These effects are expected to be relevant for Y1, which exhibits both low metallicity and low dust mass.} 
\co{Theoretical studies have shown that high dust temperatures can be achieved in compact star-forming regions \citep[e.g.,][]{Nakazato2026}.} \citet{Kano2026} suggest that the high dust temperature in Y1 may be explained by an extremely dense cold neutral medium ($\sim10^{3-4}$~cm$^{-3}$), which enhances UV photon trapping and increases the efficiency of dust heating. 
\cn{Although current observations do not indicate an elevated star-formation rate surface density for Y1 on global scales \citep{Harshan2024}, compact star-forming regions on unresolved scales may still contribute to the dust heating.}

On the other hand, we detect a broad H$\beta$ line (Section~\ref{subsec:broadline}), indicating the presence of an AGN. The spatial coincidence between the broad-line emission and one of the dust continuum peaks suggests that AGN radiation may contribute locally to dust heating. 
\citet{Kirkpatrick2012} show that AGN-host galaxies exhibit higher dust temperatures than star-forming galaxies based on stacked SED analyses. \cc{They interpret this as evidence that UV--optical radiation from the AGN is absorbed by surrounding dust and re-emitted in the infrared, producing a warm dust component in addition to that expected from star formation alone.} Figure~\ref{fig:Td_LIR} illustrates this trend, where AGN-host galaxies (orange) are offset toward higher dust temperatures by $\sim20$~K compared to star-forming galaxies (blue). 
Spatially resolved observations~\cj{\citep[e.g.,][]{Tsukui2023,Villanueva2024,Meyer2025,FernandezArada2025}} and theoretical studies~\citep[e.g., radiative transfer modeling of a quasar at $z\sim6$;][]{Schneider2015} further suggest that AGN can heat dust and contribute to the infrared emission on sub-galactic scales. \cj{In particular, the $z\sim6.9$ quasar host galaxy J2348$-$3054 exhibits a dust temperature of $88\pm2$~K within the central $r<216$~pc region \citep{FernandezArada2025}. A similar AGN-driven heating mechanism may contribute to the elevated dust temperature observed in Y1.} 
\ckt{We therefore interpret the high dust temperature of Y1 as the result of multiple contributing factors, including \cd{redshift-dependent ISM conditions~(e.g., low metallicity and low dust mass)}, compact star formation, and AGN heating}. 

\ca{Next, we consider the dust production mechanisms.} \ckt{Dust production mechanisms are broadly categorized into stellar sources (e.g., supernovae and AGB stars) and grain growth in the ISM through accretion of gas-phase metals onto pre-existing dust grains \citep{Dwek1998,Asano2013}.}
Chemical evolution models suggest that stellar sources \cd{can dominate dust production in low-metallicity galaxies, while grain growth in the ISM becomes dominant as galaxies evolve and enrich metals~\citep{Mancini2015} .}

Previous studies have shown that the efficiency of ISM grain growth is dependent on metallicity~\citep{Asano2013,Burgarella2025}. 
In particular, grain growth is expected to become the dominant source of dust production above a critical metallicity, $Z_{\rm crit}$ \citep{Asano2013,Burgarella2025}. \cb{Although $Z_{\rm crit}$ depends on model assumptions and physical conditions, such as the star formation timescale and the dust grain size distribution, it is typically estimated to be in the range of $\sim 0.1$--$0.2~Z_\odot$~\citep[e.g.,][]{Asano2013,Burgarella2025}.} Figure~\ref{fig:z_crit} shows the relation between $M_{\rm dust}$ and gas-phase metallicity, where the range of $Z_{\rm crit}$ is indicated by \cc{the gray} shaded region. \cb{Y1 has a metallicity of $Z \sim 0.15~Z_\odot$ (red diamond), placing it around the critical metallicity \cj{and in the low-metallicity regime along the sequence traced by the Cosmic Evolution Early Release Science Survey~\citep[CEERS;][]{Burgarella2025}, the ALMA survey of LABOCA Extended Chandra Deep Field South Survey sources~\citep[ALESS; $z\sim1$--6;][]{daCunha2015}, and the Reionization Era Bright Emission Line Survey~\citep[REBELS; $z\sim$7;][]{Algera2025} galaxies}.} \cd{This suggests that grain growth in the ISM is becoming efficient in Y1. }

To further investigate the dust content, we examine the dust-to-gas and dust-to-metal ratios as a function of metallicity, following \citet{Kiyota2025} (see also \citealt{Palla2024,Algera2025}), as shown in Figure~\ref{fig:Md_comp}. The molecular gas mass is estimated from the [\textsc{Cii}] luminosity using the empirical relation: 
\begin{equation}
    \log{\left( \frac{L_\mathrm{[\textsc{C\,ii}]}}{L_\odot} \right)} = -1.28 + 0.98\log{\left( \frac{M_\mathrm{H_2}}{M_\odot} \right)}
\end{equation}
~\citep{Zanella2018}. Adopting a conversion factor $M_{\rm gas} = \xi (M_{\rm HI} + M_{\rm H_2})$ with a metallicity-dependent factor of $\xi = 1.34$, and assuming $M_{\rm HI} \ll M_{\rm H_2}$, we derive a total gas mass of $M_\mathrm{gas}=5.56_{-0.61} ^{+0.61} \times 10^9~M_\odot$ \cf{, where the inferred gas mass does not include the ionized gas component}.
The total metal mass is then estimated as $M_{\rm metal} = Z~M_{\rm gas} + M_{\rm dust}$, yielding $M_\mathrm{metal}=1.24 _{-0.13} ^{+0.18} \times 10^6~M_\odot$. \cb{Here, $Z$ is the metallicity} \ca{$Z = Z/Z_\odot \times Z_\odot=10^{12+\log{(\rm O/H)}-8.69}\times Z_\odot=0.00198$, adopting a solar metallicity of $Z_\odot=0.0134$~\citep{Asplund2009}.}

From these values, we obtain \cg{$\log (M_{\rm dust}/M_{\rm gas})=-3.60^{+0.29}_{-0.22}$ and $\log (M_{\rm dust}/M_{\rm metal})=-0.95^{+0.29}_{-0.20}$}. These ratios are shown as red diamonds in Figure~\ref{fig:Md_comp}. Both ratios are lower than the sequences observed in high-$z$ samples from the REBELS~\cd{\citep[blue; $z\sim7$;][]{Algera2025}} and \cj{Quintet YD1 and YD4YD6~\citep[light blue; $z\sim7.9$][]{Umehata2025}}, as well as local galaxies \ca{\citep[black and gray (binned);][]{DeVis2019,Casasola2020}}, and submillimeter galaxies \ca{\citep[SMGs; green;][]{Shapley2020}}, indicating that Y1 is relatively dust-poor compared to its gas and metal content. These results suggest that Y1 is in an early stage of dust enrichment, consistent with grain growth becoming efficient near the critical metallicity. \cj{We note that [\textsc{Cii}] does not uniquely trace molecular gas, as it can originate from multiple ISM phases~\cq{\citep{Casavecchia2025,Vallini2025}}. In addition, its luminosity depends on metallicity (i.e., carbon abundance) and ISM conditions, introducing systematic uncertainties in the inferred gas mass.}

\begin{figure}
    \centering
    \includegraphics[width=\linewidth]{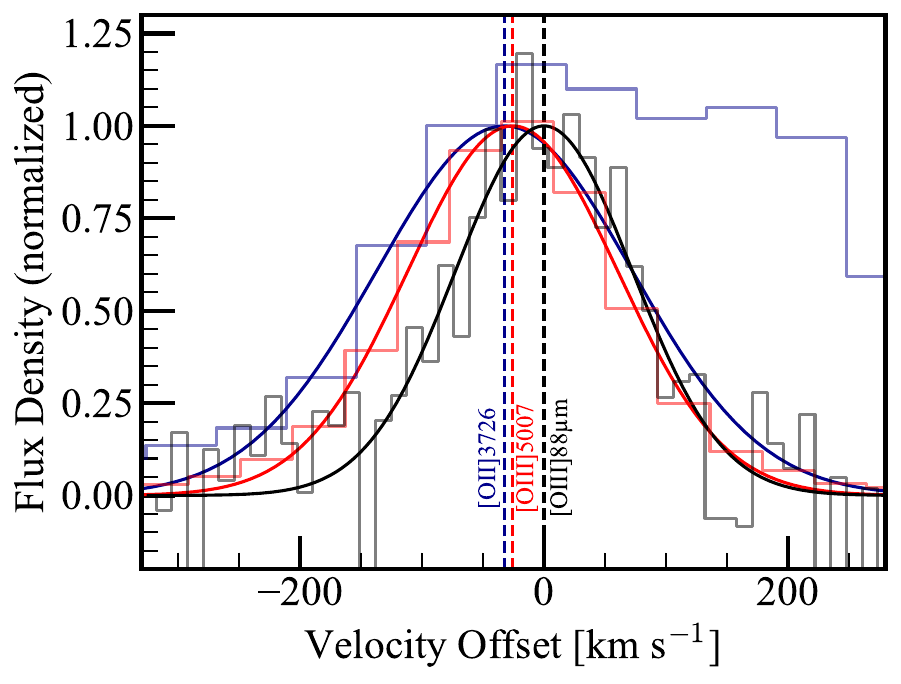}
    \caption{\ck{Velocity-offset profiles of [\textsc{Oii}]\,3726 (blue) and [\textsc{Oiii}]\,5007 (red) relative to [\textsc{Oiii}] 88$\mu$m (black; reference frame).} Solid curves show Gaussian fits to each profile. Vertical dashed lines mark the center velocity of each component.}
    \label{fig:losvel}
\end{figure}

\subsection{[\textsc{Oiii}]\,88$\mu$m/5007 Ratio}
\label{sec:o3ratdisc}

\begin{figure}
    \centering
    \includegraphics[width=.8\linewidth]{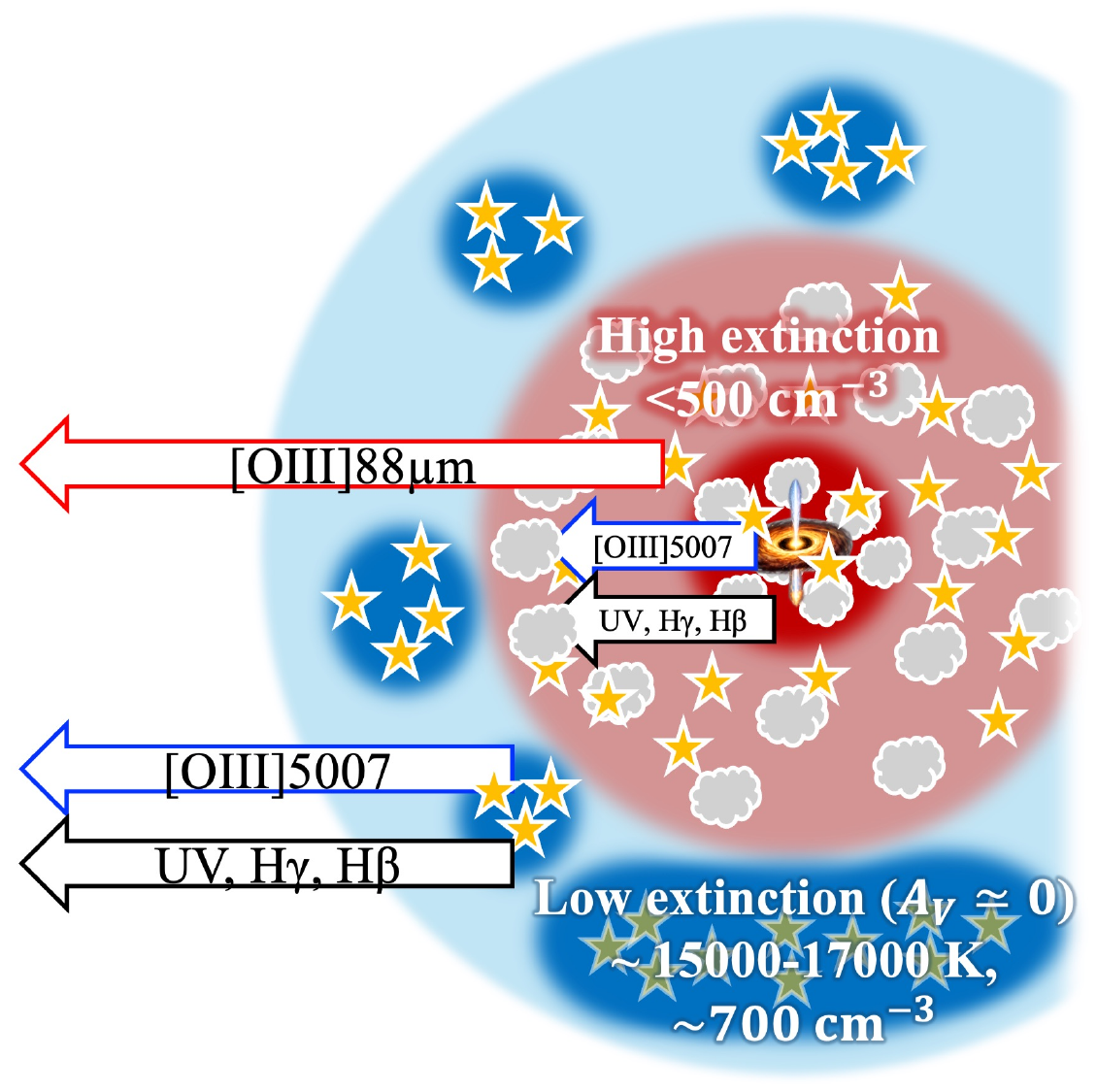}
    \caption{
    %Schematic illustration of the spatial origin of emission lines and the impact of different dust extinction. Blue, black, and red arrows denote emission traced by [\textsc{Oiii}]\,5007, UV/H$\gamma$/H$\beta$, and [\textsc{Oiii}]\,88$\mu$m, respectively. The innermost region is dense and heavily dust-obscured, surrounded by a less dense but still dusty region where [\textsc{Oiii}]\,88$\mu$m emission can originate. The outermost region has low extinction and is largely dust-free. Gray clouds denote dust, and yellow stars indicate stellar sources.
    Schematic illustration of the spatial origin of the emission lines and the effects of dust attenuation. The system consists of a dust-rich inner region and a dust-poor outer region. The innermost compact high-density region, associated with an AGN, is dust-obscured and contributes little to the observed [\textsc{Oiii}]\,88$\mu$m emission because of its high electron density. Surrounding this component is a lower-density dusty region ($n_e \lesssim 500$~cm$^{-3}$), where [\textsc{Oiii}]\,88$\mu$m emission can efficiently arise. In these dusty inner regions, dust attenuation suppresses the escape of UV and optical emission. In contrast, the outer low-extinction region dominates the observed UV continuum and optical emission lines. The outer region is characterized by low dust extinction, as suggested by the H$\gamma$/H$\beta$ ratio being consistent with the Case B value. The optical diagnostics therefore primarily trace the ionized gas in this outer component, with an electron density of $\sim700$~cm$^{-3}$ inferred from the [\textsc{Oii}] doublet ratio and an electron temperature of $T_e\sim15000\textrm{--}17000$~K inferred from the optical [\textsc{Oiii}] line ratio. Because this electron density exceeds the critical density of [\textsc{Oiii}]\,88$\mu$m, the outer low-extinction region is unlikely to contribute significantly to the observed [\textsc{Oiii}]\,88$\mu$m emission. Blue, black, and red arrows denote [\textsc{Oiii}]\,5007, UV/H$\gamma$/H$\beta$, and [\textsc{Oiii}]\,88$\mu$m emission, respectively. Gray clouds denote dust, and yellow stars indicate stellar sources. }
    \label{fig:picture}
\end{figure}

As discussed in Section~\ref{sec:o3rat}, Y1 has a high integrated flux ratio of [\textsc{Oiii}]\,88$\mu$m/[\textsc{Oiii}]\,5007$ = 0.26 \pm 0.06$. Under the assumption of a single homogeneous ionized nebula, this ratio cannot be reproduced for any electron density.
In contrast, the electron density inferred from the [\textsc{Oii}]\,3726,3729 doublet is $\gtrsim700$~cm$^{-3}$, resulting in an inconsistency between the densities inferred from [\textsc{Oii}] and [\textsc{Oiii}] diagnostics. Similar discrepancies have been reported in other high-redshift galaxies \citep{Harikane2025,Usui2025}, suggesting that this may be common in \cc{high-$z$} ISM environments~\cq{\citep{Choustikov2026}}.

One possible explanation is that the [\textsc{Oiii}]\,88$\mu$m and optical emission lines originate from physically and kinematically distinct gas components. 
To investigate this possibility, we compare the line-of-sight velocities of [\textsc{Oii}]\,3726, [\textsc{Oiii}]\,5007, and [\textsc{Oiii}]\,88$\mu$m. Figure~\ref{fig:losvel} shows the velocity offsets of the rest-frame optical lines relative to the [\textsc{Oiii}]\,88$\mu$m line. The [\textsc{Oii}]\,3726 and [\textsc{Oiii}]\,5007 spectra are obtained from the IFU-integrated spectrum, and the [\textsc{Oiii}]\,88$\mu$m spectrum is constructed by integrating all spaxels with S/N $>3$ in the moment 0 map over the same aperture used for IFU-integrated spectra. According to the \textit{JWST} User Documentation, IFU data have the wavelength calibration uncertainty of $\sim 20~\mathrm{km~s^{-1}}$~\ca{\footnote{\url{https://jwst-docs.stsci.edu/jwst-calibration-status/nirspec-calibration-status/nirspec-ifu-calibration-status\#gsc.tab=0}}}. Accounting for this uncertainty, we find velocity offsets of \ck{$-32^{+37}_{-37}~\mathrm{km~s^{-1}}$} for  [\textsc{Oii}]\,3726 and $-26^{+30}_{-30}~\mathrm{km~s^{-1}}$ for [\textsc{Oiii}]\,5007 relative to [\textsc{Oiii}]\,88$\mu$m. These offsets are consistent with zero within the uncertainties, indicating that there is no significant velocity separation between the FIR and optical line emitting regions. This suggests that the observed discrepancy is unlikely to arise from kinematically distinct components, but instead reflects different physical conditions within spatially connected regions. 

\cg{Dust attenuation is required to explain the discrepancy between the observed and theoretical [\textsc{Oiii}] line ratios.} If the electron density inferred from [\textsc{Oiii}] is assumed to be consistent with that derived from the [\textsc{Oii}]\,3726,3729 doublet, then extinction is estimated to be $A_V \sim 1.7$, \ck{adopting the Small Magellanic Cloud (SMC) reddening law \citep{Gordon2003}.} However, the integrated Balmer decrement measured from the IFU data, \cc{${\rm H}\gamma/{\rm H}\beta = 0.47^{+0.05}_{-0.04}$}, indicates little or no dust attenuation assuming Case B recombination. 
The fluxes used to derive this ratio are measured from the spectrum extracted after masking spaxels affected by the wavelength gap, using the same masked region described in Section~\ref{sec:zte}. 
\cj{The intrinsic ratio calculated using \texttt{Pyneb} is H$\gamma$/H$\beta$ = 0.47 for Case B conditions with $T_{\mathrm e} = 17300$ K and $n_{\mathrm e} = 100\textrm{--}1000$ cm$^{-3}$.} This apparent inconsistency among the multi-wavelength line diagnostics suggests that the observed optical emission lines preferentially trace low-extinction regions.

\cd{Figure~\ref{fig:picture} presents a schematic illustration of the ISM structure that can explain these observations.} \ckt{The system consists of a dust-rich inner region and a dust-poor outer region.} \cd{The center of the dust-rich region hosts a compact, high-density region associated with an AGN. Because the critical density of [\textsc{Oiii}]\,88$\mu$m is $\sim500$~cm$^{-3}$, the innermost region is unlikely to contribute significantly to the observed [\textsc{Oiii}]\,88$\mu$m emission. Surrounding this central component is a lower-density dusty region with $n_e \lesssim 500$~cm$^{-3}$, where [\textsc{Oiii}]\,88$\mu$m can arise efficiently. In these dusty inner regions, dust attenuation suppresses the escape of UV and optical emission, limiting their contribution to the observed UV--optical emission.} 
The outer region is characterized by low dust extinction, as suggested by the H$\gamma$/H$\beta$ ratio being consistent with the Case B value. As a result, the UV continuum and optical emission lines are not significantly affected by dust extinction, and the observed UV--optical emission is dominated by this component. Under this interpretation, the optical diagnostics primarily reflect the physical conditions of the ionized gas in the outer low-extinction region, with an electron density of $\sim700$~cm$^{-3}$ inferred from the [\textsc{Oii}] doublet ratio and an electron temperature of $T_e \sim15000\textrm{--}17000$~K inferred from the optical [\textsc{Oiii}] line ratio. Because this electron density exceeds the critical density of [\textsc{Oiii}]\,88$\mu$m, the outer low-extinction region is unlikely to contribute significantly to the observed [\textsc{Oiii}]\,88$\mu$m emission.

\cq{This picture is broadly consistent with observations of other high-redshift galaxies. 
The right panels of Figure~\ref{fig:o3rat_ne} show the observed [\textsc{Oiii}]\,88$\mu$m/[\textsc{Oiii}]\,5007 ratios and the theoretical relations between [\textsc{Oiii}]\,88$\mu$m/[\textsc{Oiii}]\,5007 and electron density based on the measured electron temperatures for ID4590~\citep[$z\sim8.5$;][]{Fujimoto2024}, J0217-0208~\citep[$z\sim6.2$;][]{Harikane2025}, SXDF-NB1106-2~\citep[$z\sim7.2$;][]{Harikane2025}, and  COS-2987~\citep[$z\sim6.8$;][]{Usui2025}. ID4590, J0217-0208, and SXDF-NB1106-2, shown in the top left, top right, and bottom left panels, respectively, do not exhibit the discrepancy seen in Y1, as there exists an electron density consistent with the observed [\textsc{Oiii}]\,88$\mu$m/[\textsc{Oiii}]\,5007 ratio under the measured electron temperature. Dust continuum emission has not been detected in these galaxies~\citep{Fujimoto2024,Inoue2016,Ren2023,Harikane2025}, consistent with our interpretation that they do not contain sufficiently dust-obscured regions to strongly suppress the optical emission lines and produce the observed discrepancy. In contrast, COS-2987, shown in the bottom right panel, exhibits a discrepancy between the observed and theoretical line ratios similar to that seen in Y1~\citep{Usui2025}. Although dust continuum emission has not been detected in this galaxy~\citep{Smit2018,Witstok2022}, the nondetection is still consistent with our interpretation. Assuming a dust mass comparable to that of Y1 ($M_{\rm dust}\sim10^6~M_\odot$), the dust continuum flux density in COS-2987 is expected to be below 1/4 of the current observational limit~\citep[$M_{\rm dust}\lesssim4\times10^6~M_\odot$ assuming $T_{\rm dust}=50$~K and $\beta_{\rm IR}=1.5$;][]{Witstok2022}. Therefore, COS-2987 may still host dusty regions comparable to those inferred for Y1, where the optical emission lines are significantly suppressed.}

\cq{Our findings and comparison with other high-$z$ galaxies support a picture in which dust extinction causes the observed optical and FIR emission to be dominated by different regions within the galaxy. As a result, optical and FIR lines do not necessarily trace the same ionized gas components, leading to discrepancies between optical line diagnostics and diagnostics combining optical and FIR lines. This interpretation is qualitatively consistent with studies suggesting that FIR-emitting regions are more compact than UV-emitting regions in high-redshift galaxies~\citep[e.g.,][]{Fujimoto2017,Rujopakarn2016,Boogaard2024,Kiyota2026}. Combining \textit{JWST} rest-frame optical spectroscopy with ALMA FIR line observations therefore requires careful interpretation, particularly when deriving global ISM properties under single-zone assumptions.}

\section{Summary}
\label{sec:6}
We have presented a comprehensive multi-wavelength analysis of MACS0416-Y1, a galaxy at $z = 8.312$ with the highest-redshift ALMA dust continuum detection to date. By combining deep \textit{JWST}/NIRSpec MSA spectroscopy with archival \textit{JWST}/\ca{NIRSpec} IFU and ALMA data, we investigate the AGN activity, dust properties, and ISM conditions of Y1. Our main findings are summarized as follows.

\begin{enumerate}
\item The deep NIRSpec spectrum reveals a broad H$\beta$ component with a velocity width of $\sim 1100$ km~s$^{-1}$, indicating the presence of a broad-line AGN. In addition, all clumps lie in the AGN regime of the MEx diagram, consistent with high ionization conditions across the clumps, potentially associated with AGN activity. We also perform a Little Red Dot test using the NIRCam photometry and find that none of the individual clumps or the integrated photometry of Y1 satisfy the LRD selection criteria.

\item Using the direct method, we derive a gas-phase metallicity of $12 + \log({\rm O/H}) = 7.86^{+0.09}_{-0.08}$ ($\sim 0.15~Z\odot$). Strong-line metallicity estimate based on the $R3$ index is consistent with this result. The metallicity places Y1 near the critical metallicity ($\sim 0.1$–$0.2~Z_\odot$), where dust grain growth in the ISM is becoming efficient.
%We derive the gas-phase metallicity using the direct method, obtaining $12 + \log({\rm O/H}) = 7.86^{+0.09}_{-0.08}$ ($\sim 0.15~Z\odot$). This moderately low metallicity places Y1 near the critical metallicity ($\sim 0.1$–$0.2~Z_\odot$), where grain growth in the ISM becomes efficient. 

\item Combining the metallicity, ALMA dust continuum measurements, and the gas mass inferred from [\textsc{Cii}], we derive low dust mass ratios of $\log(M_{\rm dust}/M_{\rm gas})=-3.60^{+0.29}_{-0.22}$ and $\log(M_{\rm dust}/M_{\rm metal})=-0.95^{+0.29}_{-0.20}$. These values indicate that Y1 is relatively dust-poor despite its dust continuum detection, consistent with an early stage of dust enrichment where grain growth is becoming efficient.
%suggesting that the system is in an early stage of dust enrichment where grain growth has only recently become efficient.
%\ckt{\item The dust mass reported in the literature ($M_{\rm dust} \sim 1.4 \times 10^6~M_\odot$) and low dust-to-gas and dust-to-metal ratios indicate that Y1 is relatively dust-poor compared to its gas and metal content. These results suggest that the system is in an early phase of dust enrichment, where grain growth has recently become effective.}

\item The high dust temperature of $T_{\rm dust} \simeq 91^{+62}_{-35}$ K may be driven by multiple factors, including low metallicity and dust mass, compact star formation, and AGN heating. Previous studies have shown that AGN can produce elevated dust temperature. The spatial coincidence between the broad line emission and one of the dust continuum peaks is consistent with such AGN-driven dust heating in Y1,  suggesting that AGN may contribute to the high dust temperature in Y1.

%\item The dust temperature reported in the literature ($T_{\rm dust} \simeq 91^{+62}_{-35}$ K) is significantly higher than typical values expected at $z \sim 8$. \cd{We interpret this as the result of multiple factors, including low metallicity and dust mass, and heating from AGN radiation.}

\item We find a high integrated line ratio of [\textsc{Oiii}] 88$\mu$m/[\textsc{Oiii}]\,$5007 = 0.26 \pm 0.06$, which cannot be reproduced by a single homogeneous ionized nebula model for any electron density. 
Explaining this ratio requires dust attenuation, whereas the Balmer decrement indicates little or no attenuation in the observed optical emission. We propose that dust attenuation causes the observed optical and FIR lines to preferentially trace different regions within the galaxy, with observed optical emission dominated by low-extinction regions and FIR emission tracing more obscured regions. This implies that joint analysis of \textit{JWST} optical spectroscopy and ALMA FIR line observations should be treated with caution, particularly when assuming single-zone ISM.
%We interpret this discrepancy as evidence that dust attenuation 
%We argue that dust attenuation causes the optical and FIR lines to trace physically connected but distinct regions within the galaxy, where optical emission preferentially escapes from dust-poor outer regions while FIR emission originates from more obscured dusty gas.
%\item We find a high integrated line ratio of [\textsc{Oiii}] 88$\mu$m/[\textsc{Oiii}]\,$5007 = 0.26 \pm 0.06$, which cannot be reproduced by single-zone ionized nebula models at any electron density. We propose that this discrepancy arises from a multi-phase ISM structure, where FIR emission originates from dusty, moderately dense regions, while optical emission lines preferentially trace less obscured outer regions.

\end{enumerate}

\begin{acknowledgments}
\ck{We thank Anishya Harshan, Rodrigo Herrera-Camus, Hanae Inami, Yuta Kageura, Minami Nakane, Koki Otaki, Raffaella Schneider, John Silverman, Joris Witstok, and Hiroto Yanagisawa for valuable discussions. }
\cn{This paper makes use of the following ALMA data: ADS/JAO.ALMA\#2016.1.00117.S, \#2017.1.0025.S, \#2017.1.00486.S, and \#2018.1.01241.S. ALMA is a partnership of ESO (representing its member states), NSF (USA) and NINS (Japan), together with NRC (Canada), NSTC and ASIAA (Taiwan), and KASI (Republic of Korea), in cooperation with the Republic of Chile. The Joint ALMA Observatory is operated by ESO, AUI/NRAO and NAOJ. 
This work is based in part on observations made with the NASA/ESA/CSA James Webb Space Telescope. The data were obtained from the Mikulski Archive for Space Telescopes at the Space Telescope Science Institute, which is operated by the Association of Universities for Research in Astronomy, Inc., under NASA contract NAS 5-03127 for JWST. These observations are associated with program GTO-1176 (PEARLS), GTO-1208 (CANUCS), and GO-4750 (DREAMS). The authors acknowledge the PEARLS and CANUCS teams, led by R. Windhorst  and C. J. Willott, respectively,  for developing their observational programs. The authors also acknowledge the TEMPLATES team, led by J. R. Rigdy, for establishing the \textit{JWST}/NIRSpec IFU reduction pipeline. Data used in this paper can be found in MAST at \dataset[10.17909/g8nk-0y53]{http://dx.doi.org/10.17909/g8nk-0y53}. ChatGPT was used only for language refinement.} 
\cn{This publication is based upon work supported by the World Premier International Research Center Initiative (WPI Initiative), MEXT, Japan, KAKENHI grant Nos. 25H00674 (M.O.), 26KJ1232 (T.K.), 23H00131 (A.K.I.), 21H04489, 26H02061 (H.Y.), and 22H04939 (M.H.) through the Japan Society for the Promotion of Science, and JST FOREST Program grant No. JP-MJFR202Z (H.Y.). This work was supported by the joint research program of the Institute for Cosmic Ray Research (ICRR), University of Tokyo. 
Y.N. acknowledges Flatiron Research Fellowhship. The Flatiron Institute is a division of the Simons Foundation.}

\end{acknowledgments}

\facilities{\textit{JWST}, ALMA, \textit{HST}}

%% This command is needed to show the entire author+affiliation list when
%% the collaboration and author truncation commands are used.  It has to
%% go at the end of the manuscript.
%\allauthors

%% Include this line if you are using the \added, \replaced, \deleted
%% commands to see a summary list of all changes at the end of the article.
%\listofchanges

%\appendix

%\section{}

%l
\end{document}